\PassOptionsToPackage{numbers,sort&compress}{natbib}
\documentclass[preprintnumbers,amsmath,amssymb, prd,notitlepage,nofootinbib,superscriptaddress, showkeys,twocolumn]{revtex4-2}

\usepackage{amsfonts,amssymb,amsmath}

\usepackage{graphicx}
\usepackage{multirow}
\usepackage{array}
\usepackage[utf8]{inputenc}

\usepackage{aas_macros}
\usepackage{xcolor}

\definecolor{xlinkcolor}{cmyk}{1,1,0,0}

\usepackage{graphicx}	

\usepackage{multirow}

\usepackage{bm}		    

\usepackage{siunitx}    
\DeclareSIUnit\parsec{pc}
\DeclareSIUnit\arcmin{$^\prime$}

\usepackage{bm}

\usepackage{orcidlink}
\usepackage{booktabs} 
\usepackage{makecell}
\usepackage[normalem]{ulem}

\hypersetup{colorlinks=true,linkcolor=xlinkcolor,citecolor=xlinkcolor,urlcolor=xlinkcolor}

\newcommand{\greaterthanapprox}{\mathrel{\vcenter{
  \offinterlineskip\halign{\hfil$##$\cr
    >\cr\noalign{\kern2pt}\sim\cr\noalign{\kern-2pt}}}}}
    
    \newcommand{\lessthanapprox}{\mathrel{\vcenter{
  \offinterlineskip\halign{\hfil$##$\cr
    <\cr\noalign{\kern2pt}\sim\cr\noalign{\kern-2pt}}}}}
    
\newcommand{\nv}{\hat{\bf n}}

\newcommand{\be}{\begin{equation}}        
\newcommand{\ee}{\end{equation}}

\DeclareRobustCommand{\VAN}[3]{#2}
\let\VANthebibliography\thebibliography
\def\thebibliography{\DeclareRobustCommand{\VAN}[3]{##3}\VANthebibliography}

\newcommand{\iac}[1]{{{{#1}}}}

\begin{document}
\title{Impact of Galactic non-Gaussian foregrounds on CMB lensing measurements}

\author{Irene Abril-Cabezas\orcidlink{0000-0003-3230-4589}}
\email{ia404@cam.ac.uk}
\affiliation{DAMTP, Centre for Mathematical Sciences, University of Cambridge, Wilberforce Road, Cambridge CB3 0WA, UK
}
\affiliation{Kavli Institute for Cosmology Cambridge, Madingley Road, Cambridge, CB3 0HA, UK}

\author{Frank J. Qu\orcidlink{0000-0001-7805-1068}}
\affiliation{Kavli Institute for Particle Astrophysics and Cosmology, Stanford University, 452 Lomita Mall, Stanford, CA, 94305, USA}
\affiliation{Department of Physics, Stanford University, 382 Via Pueblo Mall, Stanford, CA, 94305, USA}
\affiliation{DAMTP, Centre for Mathematical Sciences, University of Cambridge, Wilberforce Road, Cambridge CB3 0WA, UK
}
\affiliation{Kavli Institute for Cosmology Cambridge, Madingley Road, Cambridge, CB3 0HA, UK}

\author{Blake D. Sherwin\orcidlink{0000-0002-4598-9719}}
\affiliation{DAMTP, Centre for Mathematical Sciences, University of Cambridge, Wilberforce Road, Cambridge CB3 0WA, UK
}
\affiliation{Kavli Institute for Cosmology Cambridge, Madingley Road, Cambridge, CB3 0HA, UK}

\author{Alexander van Engelen}
\affiliation{School of Earth and Space Exploration, Arizona State University, Tempe, AZ 85287, USA}

\author{Niall MacCrann\orcidlink{0000-0002-8998-3909}}
\affiliation{DAMTP, Centre for Mathematical Sciences, University of Cambridge, Wilberforce Road, Cambridge CB3 0WA, UK
}
\affiliation{Kavli Institute for Cosmology Cambridge, Madingley Road, Cambridge, CB3 0HA, UK}

\author{Carlos Herv\'ias-Caimapo\orcidlink{0000-0002-4765-3426}}
\affiliation{Instituto de Astrof\'isica and Centro de Astro-Ingenier\'ia, Facultad de F\'isica, Pontificia Universidad Cat\'olica de Chile, Av. Vicu\~na Mackenna 4860, 7820436, Macul, Santiago, Chile}

\author{Omar Darwish\orcidlink{0000-0003-2946-1866}}
\affiliation{Universit\'e de Gen\`eve, D\'epartement de Physique Th\'eorique et CAP, 24 Quai Ansermet, CH-1211 Gen\`eve 4, Switzerland}

\author{J. Colin Hill\orcidlink{0000-0002-9539-0835}}
\affiliation{Department of Physics, Columbia University, New York, NY 10027, USA}

\author{Mathew S. Madhavacheril\orcidlink{0000-0001-6740-5350}}
\affiliation{Department of Physics and Astronomy, University of Pennsylvania, 209 South 33rd Street, Philadelphia, PA, USA 19104}%

\author{Neelima Sehgal\orcidlink{0000-0002-9674-4527}}
\affiliation{Physics and Astronomy Department, Stony Brook University, Stony Brook, New York 11794, USA}

\date{\today}

\begin{abstract}
Weak gravitational lensing of the CMB has been established as a robust and powerful observable for precision cosmology. However, the impact of Galactic foregrounds, which has been studied less extensively than many other potential systematics, could in principle pose a problem for CMB lensing measurements. These foregrounds are inherently non-Gaussian and hence might mimic the characteristic signal that lensing estimators are designed to measure. We present an analysis that quantifies the level of contamination from Galactic dust in lensing measurements, focusing particularly on measurements with the Atacama Cosmology Telescope and the Simons Observatory. We employ a whole suite of foreground models and study the contamination of lensing measurements with both individual frequency channels and multifrequency combinations. We test the sensitivity of different estimators to the level of foreground non-Gaussianity, and the dependence on sky fraction and multipole range used. We find that Galactic foregrounds do not present a problem for the Atacama Cosmology Telescope experiment (the bias in the inferred CMB lensing power spectrum amplitude remains below 0.3$\sigma$). For Simons Observatory, not all foreground models remain below this threshold. Although our results are conservative upper limits, they suggest that further work on characterizing dust biases and determining the impact of mitigation methods is well motivated, especially for the largest sky fractions.\medskip
\end{abstract}

\maketitle

\section{Introduction}\label{sec:intro}

The cosmic microwave background (CMB) was released when the Universe was approximately \num[group-separator = {,}]{380000} years old. As CMB photons travel to our observation point, they encounter large-scale structure that gravitationally deflects their trajectories. This phenomenon is known as weak gravitational lensing of the CMB. A measurement of this effect provides an opportunity to map the projected matter density field, probe the growth of structure, constrain the neutrino mass, learn about the geometry of our Universe, or constrain the dark energy model \citep{Blanchard_1987, Lewis_2006}.\medskip

Lensing breaks the statistical isotropy of the CMB, which in turn can be used to reconstruct the lensing deflections using quadratic combinations of the CMB intensity ($T$) and polarization ($E, B$) fields~\citep{Hu_2001, Hu_2002, Okamoto_2003}. In particular, different quadratic estimators (QEs) provide measurements with varying signal-to-noise ratios (SNRs) depending on the experiment's noise levels. The $TT$ estimator has the highest SNR for an experiment such as \textit{Planck} (with noise levels around $\SI{30}{\micro\kelvin}$-arcmin in temperature) \citep{Planck_2020_mission, Planck_lensing_2020, Carron_2022_PR4lensing}. Moreover, the reconstructed deflection field is still dominated by temperature-only information with current CMB experiments such as the Atacama Cosmology Telescope \citep[ACT,][]{Fowler_2007, Thronton_2016, Henderson_2016, Frank_ACT_lensing_2024} or the Simons Observatory \citep[SO,][]{SO_2019}, all with temperature noise levels at least three times lower than \textit{Planck}. The $EB$ estimator becomes the most powerful and will dominate the lensing reconstruction once experiments reach noise levels well below $\SI{5}{\micro\kelvin}$-arcmin \citep{Zaldarriaga_1998, Hu_2002}.\footnote{CMB temperature maps again become the dominant contributor to the SNR for a CMB experiment such as CMB-HD \citep{CMB_HD, Nguyen_2019}.} Next-generation experiments, including CMB Stage IV \citep{Abitbol_2017,Barron_2022},  will most likely not use quadratic estimators to reconstruct the lensing field (as they will become suboptimal), and will instead favor likelihood-based iterative methods \citep{Hirata_2003, Carron_MAP_2017, Legrand_2022, Legrand_2023} (as recently used by the South Pole Telescope \citep[SPT,][]{Benson_2014, Pan_SPT3G_2023_lensing}). Nevertheless, QEs (and in particular $TT$) will remain a useful tool to forecast and cross-check results. \medskip 

Alongside lensing, foregrounds are another potential source of statistical anisotropy and non-Gaussianity in CMB maps. This is particularly important in temperature, for which extragalactic foregrounds such as the thermal Sunyaev-Zel'dovich effect \cite{Zeldovich_1969, Sunyaev_1970} sourced by galaxy clusters or the millimeter-wave emission by dusty star-forming galaxies \citep{Puget_1996, Dwek_1998, Fixsen_1998} can significantly bias CMB lensing measurements, as multiple studies \citep[e.g.][]{Osborne_2014, vanEngelen_2014, Madhavacheril_2018, Omar_2021, MacCrann_2024, Doohan_2025} have shown. A common mitigation strategy in this case is to use ``bias-hardened'' quadratic estimators, a modified form of the standard QE that has zero response, at leading order, to the statistical anisotropy caused by these sources \citep{Namikawa_2013, Osborne_2014, Sailer_2020}. In polarization, extragalactic foreground biases are not expected to be significant \citep{Smith_2009, Hall_2014, Feng_2020}. Moreover, new methods are being developed to further minimize their impact \citep{Namikawa_2014, Sailer_2023, Qu_2024_pol}. \medskip

Galactic foregrounds, on the other hand, have not been studied to the same extent in the CMB lensing literature (with only a few papers discussing the problem, e.g.~\citep{Fantaye_2012, Beck_2020, Challinor_2018}). Nevertheless, they are of particular concern for lensing studies because they are highly non-Gaussian{, with interstellar medium (ISM) structures correlating different scales in a highly anisotropic fashion,} and hence could mimic the characteristic signal that lensing estimators are designed to measure. Thermal dust emission from our own Galaxy, in particular, is the most relevant Galactic emission mechanism at frequencies above $\nu \gtrsim \SI{70}{\giga\hertz}$ \cite[see e.g.][]{Planck_2015_dust, Plack_2016_X, Planck_2020_fgs, Planck_2020_XI}. In polarization, \cite{Fantaye_2012, Beck_2020} find using one non-Gaussian small-scale dust realization that dust produces non-negligible biases, and suggest parametric foreground cleaning approaches that successfully mitigate them. \cite{Challinor_2018} performs an approximate estimate of dust biases using a small high-dust region near the Galactic plane and finds a \SI{15}{\percent} bias in the temperature reconstruction, which is reduced to acceptable levels after foreground cleaning; the authors note, however, that some caution is required in interpreting this result more broadly since the these are specific to the bright dust field they employ.\medskip

Compared to \emph{Planck}, ground-based experiments have a reduced frequency coverage which hampers component-separation techniques.\footnote{This will soon change with the upcoming Simons Observatory, operating on six frequency bands centered around \SI{27}{}, \SI{39}{}, \SI{93}{}, \SI{145}{}, \SI{225}{} and \SI{280}{\giga\hertz} \citep{SO_2019}.} This results in different strategies to minimize foreground contamination. While \emph{Planck} used foreground-cleaned CMB maps \citep{Planck_lensing_2020, Carron_2022_PR4lensing}, lensing reconstruction from ground-based experiments such as ACT \citep{Frank_ACT_lensing_2024} or SPT \citep{Pan_SPT3G_2023_lensing} only use CMB maps from at most three frequency bands, centered at \SI{90}{}, \SI{150}{} and \SI{220}{\giga\hertz}. These analyses performed a whole suite of null tests that ensured minimal contamination from extragalactic foregrounds \citep[e.g.][]{MacCrann_2024}. Since detailed estimates of Galactic foreground biases were not available, sky fractions were chosen very conservatively, potentially discarding a large fraction of usable data. A detailed quantification of the level of bias on these experiments is hence well motivated.\medskip 

In this work, we focus on the impact of Galactic dust non-Gaussianity on lensing reconstructions performed with quadratic estimators. Estimating the level of bias has been difficult in the past because, as pointed out in \cite{Challinor_2018}, the dust field on small scales is not well constrained by current data, especially in polarization and at high Galactic latitudes where the dust emission is reduced. When simulating the foreground field, this issue is usually circumvented by filtering out these scales and replacing them with Gaussian realizations of a given target power spectrum that follows the trend measured on large scales \citep[see e.g.][]{Hervias_2016, Thorne_2017}. However, this solution does not account correctly for non-Gaussianity on small scales. Fortunately, several groups have produced foreground simulations including non-Gaussianities on small scales in the last few years. These include magneto-hydrodynamic simulations \citep[e.g.][]{Kritsuk_2018,Kim_2019}, phenomenological models based on temperature observations using a superposition of Galactic dust layers \citep[e.g.][]{Vansyngel_2017, Martinez_2018}, models built from wavelet scattering transform statistics \citep{Allys_2019, Regaldo_2020, Mousset_2024} or generative adversarial neural networks \citep{Krachmalnicoff_2021, Yao_2024}, dust models based on neutral hydrogen \citep{Clark_2019}, and filament-based ones \citep{huffenberger_2020, DF_2022}. In particular, the \textsc{PySM} software \citep{Thorne_2017, pysm} provides a suite of models spanning a range of complexities on the scales where there is little data available, hence making it an ideal tool to check for Galactic foreground biases in a variety of physical scenarios regarding the foreground properties. In this work, we will use a whole range of simulations to estimate the bias to the lensing power spectrum due to Galactic foregrounds. {Previous studies in the literature have focused on previous \citep{Perotto_2010} or longer-term future CMB missions \citep{Fantaye_2012, Challinor_2018, Beck_2020}. Instead, we will explore the impact it has on lensing from current generation experiments, in both temperature and polarization, and for a whole suite of dust models.}\medskip

The paper is organized as follows. We present the general methodology of lensing reconstruction in $\S$\ref{sec:methods}, with a particular focus on the approach taken in this work. In $\S$\ref{sec:results}, we present upper limits to the level of Galactic dust contamination on CMB lensing reconstruction analyses and their bias to the lensing amplitude parameter. We conclude in $\S$\ref{sec:conclusion}. We assume the following default cosmology: $H_0 = \SI{67.02}{\kilo\metre\per\second\per\mega\parsec}$, $\Omega_b h^2 = 0.0222 $, $\Omega_c h^2 = 0.1203$. \medskip

\section{Methodology}\label{sec:methods}

\subsection{Formalism of lensing reconstruction with quadratic estimators}

Weak lensing of the CMB leads to a remapping of the observed CMB sky, which can be expressed to leading order in $\phi$ as  \citep{Lewis_2006, Hanson_2010_review}:
\begin{equation}\label{eq:remap}
   T(\nv) = \widetilde{T}\left(\nv + \nabla\phi\left(\nv\right)\right) \simeq \widetilde{T}(\nv) + \nabla\widetilde{T}\left(\nv\right) \cdot \nabla\phi\left(\nv\right),
\end{equation}
where $T$ is the lensed temperature field and $\widetilde{T}$ is the unlensed temperature field.\footnote{Similar expressions can be derived for the polarization component of the CMB.} The lensing deflection angle $\nabla\phi$ is expressed in terms of the lensing potential $\phi$, itself an integral of the gravitational potential\footnote{More precisely, CMB lensing measurements are sensitive to the Weyl potential $\Psi_\mathrm{W} \equiv \left(\Phi+\Psi\right)/2$, where $\Psi$ and $\Phi$ are the temporal and spatial distortions of the Universe's geometry,
\begin{equation}
\textrm{d}s^2= a^2(\tau)\left[-\left(1+2\Psi\right)\textrm{d}\tau^2 + \left(1-2\Phi\right)\textrm{d}\bf{x}^2 \right].
\end{equation} The so-called anisotropic stress describes the difference between the time distortion and the spatial distortion of the metric. This is usually assumed to be negligible, in which case we have $\Psi_\mathrm{W}=\Psi=\Phi$.
} along the line of sight $\nv$. From the equation above, we see that lensing changes the statistics of the CMB, inducing correlations between the temperature field and its gradient. When averaging over all possible unlensed CMB realizations, non-opposite $\mathbf{l}$ modes in the CMB now become correlated due to a $\phi(\mathbf{L})$ lensing mode:
\begin{equation}\label{eq:TT}
    \langle T\left(\mathbf{l}\right)T(\mathbf{L} - \mathbf{l})\rangle_{\mathrm{CMB}} = f(\mathbf{l}, \mathbf{L}) \phi(\mathbf{L}), \quad \mathbf{L} \neq 0,
\end{equation}
where $f(\mathbf{l}, \mathbf{L})$ is a function that depends on the angular power spectrum of the unlensed CMB temperature.\footnote{In practice, we use the lensed gradient power spectra $C_\ell^{{X\nabla Y}}$ for $X,Y=T,E,B$ to reduce higher-order corrections needed to match the non-perturbative result \citep{Lewis_2011, Fabbian_2019}.} This mode-coupling is a reflection of the fact that lensing (with a fixed lensing field) breaks translation invariance in the CMB. Quadratic estimators \citep{Hu_2002} exploit this mode-coupling property to provide an estimate for $\phi(\mathbf{L})$:
\begin{equation}\label{eq:phihat}
     \hat{\phi}(\mathbf{L})  = A_\mathbf{L}^\phi \int \frac{\mathrm{d}^2\mathbf{l}}{\left(2\pi\right)^2} \frac{f(\mathbf{l}, \mathbf{L}) T\left(\mathbf{l}\right)T(\mathbf{L}-\mathbf{l})}{2 C_\ell^{\mathrm{obs}} C_{| \mathbf{L} - \mathbf{l}|}^{\mathrm{obs}}},
\end{equation}
where $C_\ell^{\mathrm{obs}}$ is the observed (lensed) CMB spectrum including noise. The quadratic estimator is normalized as follows: 
\begin{equation}\label{eq:anorm}
     A_\mathbf{L}^\phi  = \left[\int \frac{\mathrm{d}^2\mathbf{l}}{\left(2\pi\right)^2} \frac{f^2(\mathbf{l}, \mathbf{L}) }{2 C_\ell^{\mathrm{obs}} C_{| \mathbf{L} - \mathbf{l}|}^{\mathrm{obs}}}\right]^{-1}.
\end{equation}
It can be shown that, if run on a map containing all the expected sources of power, $A_\mathbf{L}^\phi$ is equivalent to the CMB lensing power spectrum reconstruction noise, also known as the $N^{(0)}_L$ bias \citep[][see also $\S$\ref{sec:app_QE}]{Hu_2001, Kesden_2003, Hanson_2011}. Below, we will perform lensing reconstruction on simulated foreground-only maps, in order to estimate the resulting bias to the lensing power spectrum. {In doing this, we implicitly assume that the dust and lensed CMB maps are entirely independent, such that we only need to compute the dust trispectrum to estimate the biases to our lensing spectrum measurement. In this case, the foreground-only map has power $C_\ell^{\mathrm{fg}}$, rather than $C_\ell^{\mathrm{obs}}$. In this case, the $N^{(0)}_L$ bias takes the form \citep{MacCrann_2024}: 
\begin{equation}
    N^{(0)}_L = \bigl(A_\mathbf{L}^\phi\bigr)^2 \int \frac{\mathrm{d}^2\mathbf{l}}{\left(2\pi\right)^2} \frac{f^2(\mathbf{l}, \mathbf{L}) C_\ell^{\mathrm{fg}} C_{| \mathbf{L} - \mathbf{l}|}^\mathrm{fg}}{2 \left(C_\ell^{\mathrm{obs}} C_{| \mathbf{L} - \mathbf{l}|}^{\mathrm{obs}}\right)^2}.
\end{equation}
This result is derived explicitly in $\S$\ref{sec:app_QE}. Finally, we note that one can work equivalently with the lensing potential $\phi$ or the convergence $\kappa$. They are related by:
\begin{equation}
    \kappa(\nv) = -\nabla^2\phi(\nv)/2.
\end{equation}
From Poisson's equation, $\kappa$ measures the projected fractional matter overdensity, instead of the projected gravitational potential traced by $\phi$. In terms of 2-point functions,
\begin{equation}\label{eq:clkkclpp}
    C_L^{\kappa\kappa} = \left(\frac{L\left(L+1\right)}{2}\right)^2 C_L^{\phi\phi}.
\end{equation}

The results presented in this section are derived in the flat-sky approximation to aid the reader. In practice, we use curved-sky calculations as implemented in \textsc{falafel}.\footnote{\url{https://github.com/simonsobs/falafel}} Similarly, lensing deflections can be reconstructed not only from the CMB temperature field, but also using different quadratic combinations of the CMB intensity ($T$) and polarization ($E$, $B$). 
These can be combined into a minimum variance MV estimator \citep{Okamoto_2003} to improve the signal-to-noise ratio. When only polarization-derived measurements are used, the minimum-variance estimator is denoted MVPol. We use \textsc{tempura}\footnote{\url{https://github.com/simonsobs/tempura}} to compute the normalizations of all estimators and perform lensing reconstructions with \textsc{so-lenspipe}.\footnote{\url{https://github.com/simonsobs/so-lenspipe}}

\subsection{Experimental setup}\label{sec:experimental}

We explore the impact of Galactic dust on lensing reconstruction for two different experimental configurations. Firstly, we focus on an ACT-like experiment, comprising two frequency bands centered at \SI{90}{} and \SI{150}{\giga\hertz}. The footprint for such analysis is constructed by intersecting the ACT footprint with \emph{Planck} masks,\footnote{\texttt{HFI\_Mask\_GalPlane-apo0\_2048\_R2.00.fits}, available on the Planck Legacy Archive (\url{https://pla.esac.esa.int/}). This is the most straightforward way to remove Galactic contamination.} and further removing regions with clearly visible Galactic dust clouds, as described in \citep{Frank_ACT_lensing_2024}. \textit{Planck} masks are labeled as XX\SI{}{\percent}, where XX denotes their sky fraction (in percent).  
The resulting masks are ACT + \textit{Planck} \SI{60}{\percent}, ACT + \textit{Planck} \SI{70}{\percent} and ACT + \textit{Planck} \SI{80}{\percent}. For short, we refer to them as ACT \SI{60}{\percent}, ACT \SI{70}{\percent} and ACT \SI{80}{\percent}. They have effective sky fractions $f_{\mathrm{sky}}=0.28$,  $f_{\mathrm{sky}}=0.31$,  and $f_{\mathrm{sky}}=0.34$, respectively, after apodizing the edges on \ang{3} scales with a cosine profile. These are shown in the top panel of Figure \ref{fig:planck_353_ACT_SO_masks}. \medskip

We take the Simons Observatory experiment, as described in \citep{SO_2019}, as the second experimental setup that we explore. We assume six SO frequency bandpasses, centered around $27$ and $\SI{39}{\giga\hertz}$ (low-frequency bands, LF), $93$ and $\SI{145}{\giga\hertz}$ (mid-frequency bands, MF), and $225$ and $\SI{280}{\giga\hertz}$ (ultra-high-frequency bands, UHF). We focus on the Large Aperture Telescope (LAT) survey of SO. Similar to ACT, we construct three Galactic masks for this configuration by intersecting the SO-LAT footprint\footnote{\texttt{LAT\_hits\_2048\_galcoords.fits}, available on NERSC upon request.} with \emph{Planck} $\SI{60}{\percent}$, $\SI{70}{\percent}$ and $\SI{80}{\percent}$ masks and further removing the regions flagged by ACT \citep{Frank_ACT_lensing_2024}. Their edges are also apodized on \ang{3} scales with a cosine profile, and the resulting sky fractions are $f_\mathrm{sky}=0.37$, $f_\mathrm{sky}=0.42$ and $f_\mathrm{sky}=0.47$, respectively. The masks are shown in the bottom panel of Figure \ref{fig:planck_353_ACT_SO_masks}. Similar to the labeling used for ACT masks, we refer to SO masks (for short) as SO \SI{60}{\percent}, SO \SI{70}{\percent} and SO \SI{80}{\percent}. \medskip

In this work, for the purposes of filtering and normalizing the quadratic estimator as above (equation \ref{eq:phihat}), we will make the simplifying assumption that the total power in the maps can be expressed as 
\begin{equation}\label{eq:cell}
    C^{\textrm{obs}}_\ell = C_\ell + \Delta^2 e^{\ell(\ell+1)\sigma^2_{\textrm{FWHM}}/8\ln{2}},
\end{equation}
where $C_\ell$ is the theoretical lensed power spectrum (computed using \textsc{camb} \citep{CAMB}). $\Delta$ is the white-noise level of the ground-based experiment. We consider two different white-noise levels. In temperature, $\Delta_T=\SI{12}{\micro\kelvin}_\mathrm{CMB}$-arcmin (ACT-like) or $\Delta_T=\SI{6}{\micro\kelvin}_\mathrm{CMB}$-arcmin (SO-like). Both have angular resolution $\sigma_{\textrm{FWHM}}=1.4^\prime$. In polarization, we assume that the white-noise levels are a factor $\sqrt{2}$ larger than in temperature.\footnote{Any experiment sensitive to polarization will always need $\times 2$ more observations to achieve the same level of sensitivity in temperature than in polarization, since polarization measurements require solving for two ($Q$, $U$) fields.} \medskip

\begin{figure*}
\includegraphics[width = 0.9\textwidth]{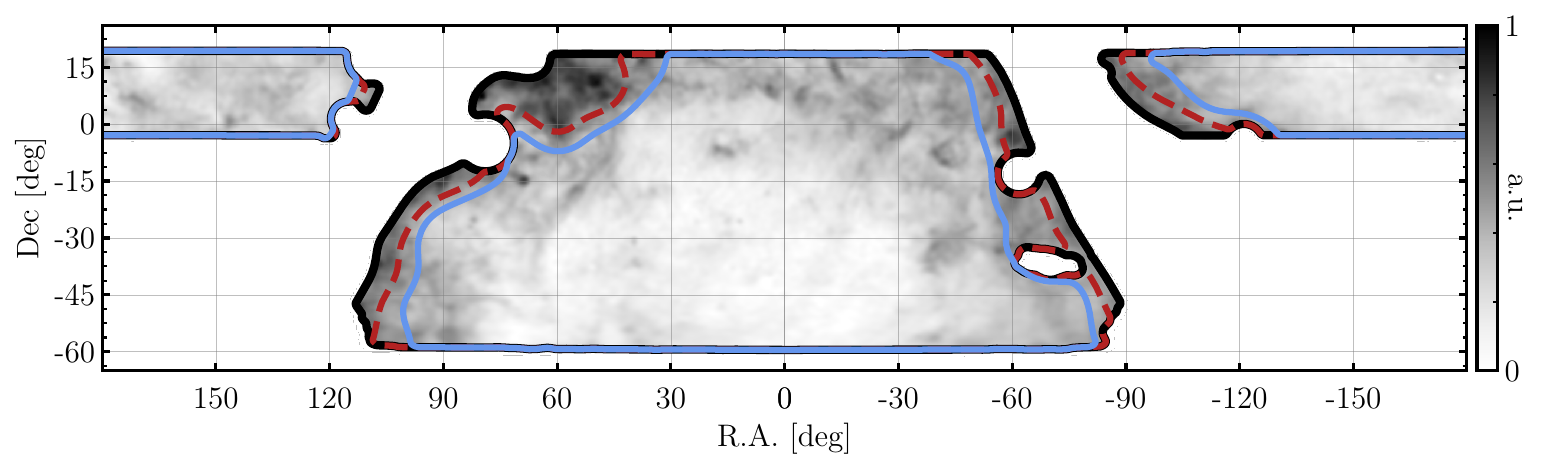}
\includegraphics[width = 0.9\textwidth]{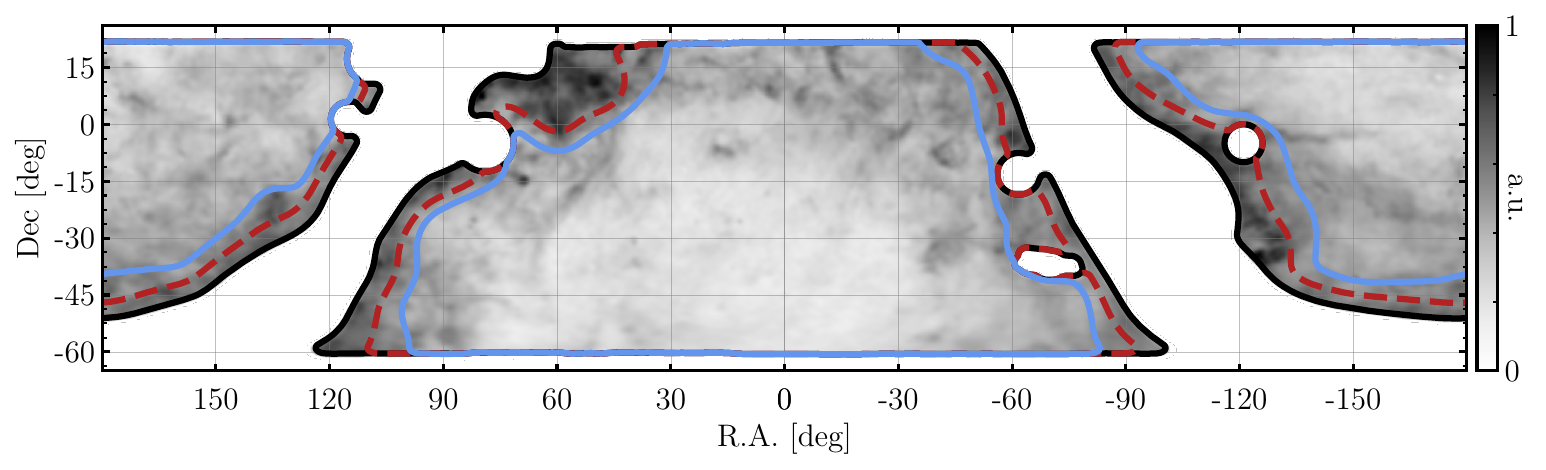}
\caption{Galactic masks for the two surveys explored in this work, projected in equatorial coordinates. They correspond to the intersection of the survey footprint with \textit{Planck} masks labeled as  XX\SI{}{\percent}, where XX denotes the sky fraction of the \textit{Planck} mask in percent. Their edges are apodized on \ang{3} scales with a cosine profile, and regions with clearly visible Galactic dust clouds are also removed, following \citep{Frank_ACT_lensing_2024, Madhavacheril_2024}. For reference, we also show \textit{Planck} \SI{353}{\giga\hertz} GNILC dust emission map \citep{Planck_2020_fgs}, where the colorbar (in arbitrary units, a.u.) indicates the strength of the emission. \textit{Top}: ACT + \textit{Planck} \SI{60}{\percent} (solid blue), ACT + \textit{Planck} \SI{70}{\percent} (dashed red) and ACT + \textit{Planck} \SI{80}{\percent} (solid black) masks, with effective sky fractions $f_{\mathrm{sky}}=0.28$,  $f_{\mathrm{sky}}=0.31$, and $f_{\mathrm{sky}}=0.34$, respectively. \textit{Bottom}: SO + \textit{Planck} \SI{60}{\percent} (solid blue) SO + \textit{Planck} \SI{70}{\percent} (dashed red) and SO + \textit{Planck} \SI{80}{\percent} (solid black) masks, with effective sky fractions $f_\mathrm{sky}=0.37$, $f_\mathrm{sky}=0.42$ and $f_\mathrm{sky}=0.47$.}
\label{fig:planck_353_ACT_SO_masks}
\end{figure*}

\subsection{Non-Gaussian Galactic sky model}\label{sec:skymodel}

Before summarizing the foreground models used in this work, we will introduce the common empirical relations employed to describe the Galactic dust component. These same relations are used when simulating the dust field. We work in thermodynamic temperature units, $\SI{}{\micro\kelvin}_\mathrm{CMB}$, for which spectral energy distribution (SED) of the CMB as function of frequency $\nu$ is a constant by definition:
\begin{equation}\label{eq:cmb_sed}
    S_\nu^{\rm{CMB}} = 1.
\end{equation}

The SED of thermal dust emission ($d$) is often modeled as a modified black-body spectrum \citep[see e.g.][]{Planck_2020_fgs}:
    \begin{equation}\label{eq:sed_dust}
        S_\nu^{d} = \frac{g(x_{\nu_0^d})}{g(x_\nu)}\left( \frac{\nu}{\nu_0^d}\right)^{\beta_d} \frac{B_\nu(\Theta_d)}{B_{\nu_0^d}(\Theta_d)},
    \end{equation}
where $\beta_d$ is the dust spectral index, $\nu_0^{d}=\SI{353}{\giga\hertz}$ is the pivot frequency, and $B_\nu(\Theta)$ is the Planck black-body spectrum, which expresses the spectral radiance of a black-body for frequency $\nu$ at temperature $\Theta$:
\begin{equation}
    B_\nu(\Theta) = \frac{2h\nu^3}{c^2}\left[\exp{\left(\frac{h\nu}{k\Theta}\right) - 1}\right]^{-1},
\end{equation}
and where $k$ and $h$ are the Boltzmann and Planck constants, respectively. $g(x_\nu)$ is the conversion factor between Rayleigh-Jeans brightness temperature units (commonly used for foregrounds) and  thermodynamic temperature units \citep{Planck_2013_units}:
    \begin{equation}
            g(x_\nu) = e^x \left(\frac{x}{e^x - 1}\right)^2, \quad x = \frac{h\nu}{k \Theta_{\rm{CMB}}},
    \end{equation}
where $\Theta_{\rm{CMB}} = \SI{2.7255}{\kelvin}$ is the CMB temperature \citep{Fixsen_2009}. \medskip

Regardless of the exact values that $\Theta_d$ and $\beta_d$ take within the different dust models, a simple evaluation of equation \eqref{eq:sed_dust} yields that dust intensity at \SI{150}{\giga\hertz} is $\sim 3$ times brighter than in the \SI{90}{\giga\hertz} channel. As a result, we expect biases in the lensing reconstruction to be almost $2$ orders of magnitude larger in the $\SI{150}{\giga\hertz}$ channel than in the $\SI{90}{\giga\hertz}$ one, since we are working with a four-point function of the dust amplitude.\medskip

Empirically, the dust ($d$) auto-spectrum is often parametrized as a power-law:
\begin{equation}
    \frac{\ell(\ell + 1)}{2\pi}C_\ell^{dd} = A_d \left(\frac{\ell}{\ell_0}\right)^{\alpha_d},
\end{equation}
where $A_d$ is the amplitude, $\alpha_d$ the tilt, and $\ell_0 = \SI{80}{}$ the pivot scale. {\textit{Planck} measures these spectra} up to {$\ell\sim 1000$} \citep{Planck_2015_dust, Plack_2016_X, Planck_2020_fgs, Planck_2020_XI}, hence the models of Galactic foreground emission that we use in this work are driven by the existing data on those scales. {\textit{Planck} small-scale measurements} are noise dominated, especially in polarization, and hence unable to inform any modeling\footnote{{Nevertheless, several works have studied different approaches to quantify the level of non-Gaussianity and statistical anisotropy of Galactic foregrounds, see e.g. \citep{Kamionkowski_2014, Rotti_2016, Philcox_2018}.}}. Different models then have different strategies to generate the (non-Gaussian) small-scales, which we now describe:\footnote{We also considered including \textsc{tigress} \citep{Tigress_2017, Tigress_2019, Tigress_2021} synthetic dust models, or those produced by \citep{Irfan_2019} or \citep{Delouis_2022}. However, they did not have enough resolution for our purposes.}
\begin{itemize}
    \item PySM \texttt{d9}. PySM models are based on \emph{Planck} dust templates \citep{Planck_2016_gnilc, Planck_2020_fgs, Planck_2020_galactic}, which are constructed using the Generalized Needlet Internal Linear Combination (GNILC) algorithm \citep{Remazeilles_2011}. The models are calibrated to observations at $\SI{353}{\giga\hertz}$ in the \emph{Planck} \texttt{GAL097} mask.\footnote{\emph{Planck} masks are labeled as GALXXX, where XXX denotes the sky fraction in percent.} PySM \texttt{d9} is the lowest complexity model provided by \cite{pysm}. It introduces non-Gaussian small-scale fluctuations in $A_d$ by modulating the small-scale emission by the large-scale signal. The spectral parameters remain uniform across the sky ($\beta_d=1.54, \Theta_d = \SI{20}{\kelvin}$), resulting in \SI{100}{\percent} correlation across frequency channels.
    \item PySM \texttt{d10}. Similar model construction as to PySM \texttt{d9}. This model presents an additional complexity, where small-scale fluctuations are present both in dust amplitude and spectral parameters. The latter leads to frequency decorrelation.
    \item PySM \texttt{d12}. PySM implementation of \cite{Martinez_2018}, where the final model is comprised of a superposition of six Galactic dust layers. The dust amplitudes on large scales follow \emph{Planck} data \citep{Planck_2016_gnilc, Planck_2020_XI} and the spectral parameters are generated randomly within each layer \citep{Martinez_2018}.
    \item \textsc{dustfilaments}\footnote{\url{https://github.com/ chervias/DustFilaments}} \citep{DF_2022, Hervias_2024}. Galactic dust exhibits a filamentary structure that cannot be reproduced from random Gaussian statistics. These filaments have been observed at millimeter wavelengths by \emph{Planck}~\citep[e.g.][]{planck_int_XXXIII,planck_int_XXXVIII}, among others, and also found to be traced by Galactic neutral hydrogen emission \citep{Clark_2014,Clark_2015, Clark_2019, Halal_2024}. Filaments and their misalignment with respect to the Galactic magnetic field can quantitatively reproduce the properties of Galactic dust as observed by \emph{Planck}~at \SI{353}{\giga\hertz} \citep{planck_int_XXX,Planck_2020_XI, huffenberger_2020}. Based on this idea, \textsc{dustfilaments}~constructs a full-sky simulation of the Galactic dust at millimeter frequencies by integrating the emission of millions of filaments in a predefined volume. By construction, the filament population is calibrated to match the dust power-spectrum on the SO-LAT footprint that intersects with \textit{Planck} 80\% Galactic mask, as described in \cite{DF_2022, Hervias_2024}, with spectral parameters derived from the GNILC best-fit $\beta_d$ and $\Theta_d$ maps \citep{Planck_2016_gnilc}.
    \item Vansyngel \texttt{d1} \citep{Vansyngel_2017}. Phenomenological model using a superposition of (seven) Galactic dust layers. Every layer follows the same intensity observations provided by \emph{Planck} GNILC maps \citep{Planck_2016_gnilc}. The maps are calibrated to match the Large Region 33 ($f_\textrm{sky}=0.33$) Galactic mask in \cite{Planck_2020_XI}, as described in \cite{Vansyngel_2017}. In temperature, there is a loss of power above $\ell\sim 1000$ due to the presence of a $30^\prime$ FWHM beam. As a result, the biases derived from this model will be slightly optimistic.    This model also introduces $EB$ correlations in the dust field at the level compatible with existing data \citep{planck_int_XXX}.
\end{itemize}

\medskip

For illustrative purposes, Figure \ref{fig:cutouts} shows $10\times10~\SI{}{\deg\squared}$ zoomed-in regions centered at $\left(\textrm{RA, DEC}\right) = \left(\SI{-42}{^\circ},\SI{-5}{^\circ}\right)$ for PySM \texttt{d9}, \textsc{dustfilaments} and Vansyngel \texttt{d1} dust emission models (in arbitrary units). They all trace the \textit{Planck} GNILC  map \citep{Planck_2016_gnilc} (upper left corner) on the largest scales, and include small-scale features unresolved with current data. In Figure \ref{fig:2pt_dusts_GAL060_90GHz}, we present the temperature power spectra at $\SI{93}{\giga\hertz}$ for the whole variety of dust models used in this work. These are computed on the SO \SI{70}{\percent} mask. Even though all models are calibrated using \SI{353}{\giga\hertz} \textit{Planck} data at the pivot scale $\ell_0=80$, we see a spread at the level of \SI{0.4}{dex} in power. This is due to both differences in the sky footprint used to calibrate their models and the spectral energy distribution assumed to extrapolate between frequencies. {In addition, the PySM models are calibrated primarily with the \textit{Planck} GNILC dust template~\citep{Planck_2016_gnilc,Planck_2020_galactic}, while \textsc{dustfilaments} is calibrated with the spectra of the \SI{353}{\giga\hertz} \textit{Planck} NPIPE maps (with the best-fit CMB subtracted). The latter case has more power, which could originate from the cosmic infrared background (CIB), as evidenced by figure~4 of~\cite{pysm}}. We do not correct for these differences when reporting the biases that different models produce.\medskip

\begin{figure}
\includegraphics[width=0.9\columnwidth]{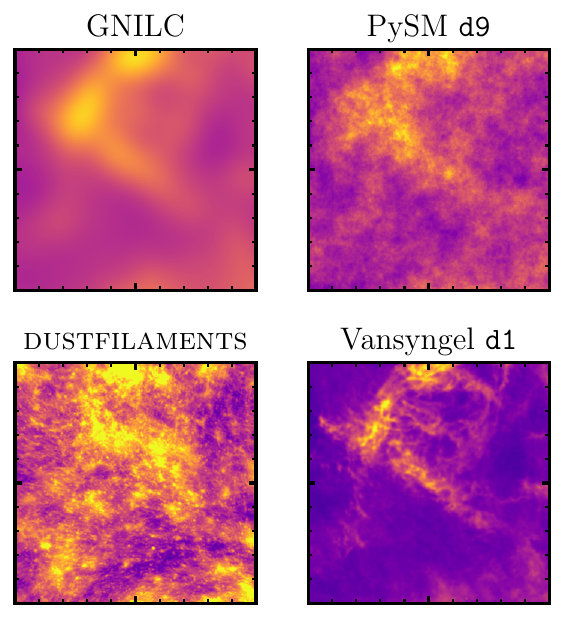}
    \caption{$10\times10~\SI{}{\deg\squared}$ zoomed-in regions centered at $\left(\textrm{RA, DEC}\right) = \left(\SI{-42}{^\circ},\SI{-5}{^\circ}\right)$ for PySM \texttt{d9}, \textsc{dustfilaments} and Vansyngel \texttt{d1} dust emission models in temperature at \SI{353}{\giga\hertz} (in arbitrary units). They all trace the \textit{Planck} GNILC  map \citep{Planck_2016_gnilc} (upper left corner) on the largest scales, and include small-scale features unresolved with current data.}
    \label{fig:cutouts}
\end{figure}

\begin{figure}
\includegraphics[width = 0.9999\columnwidth]{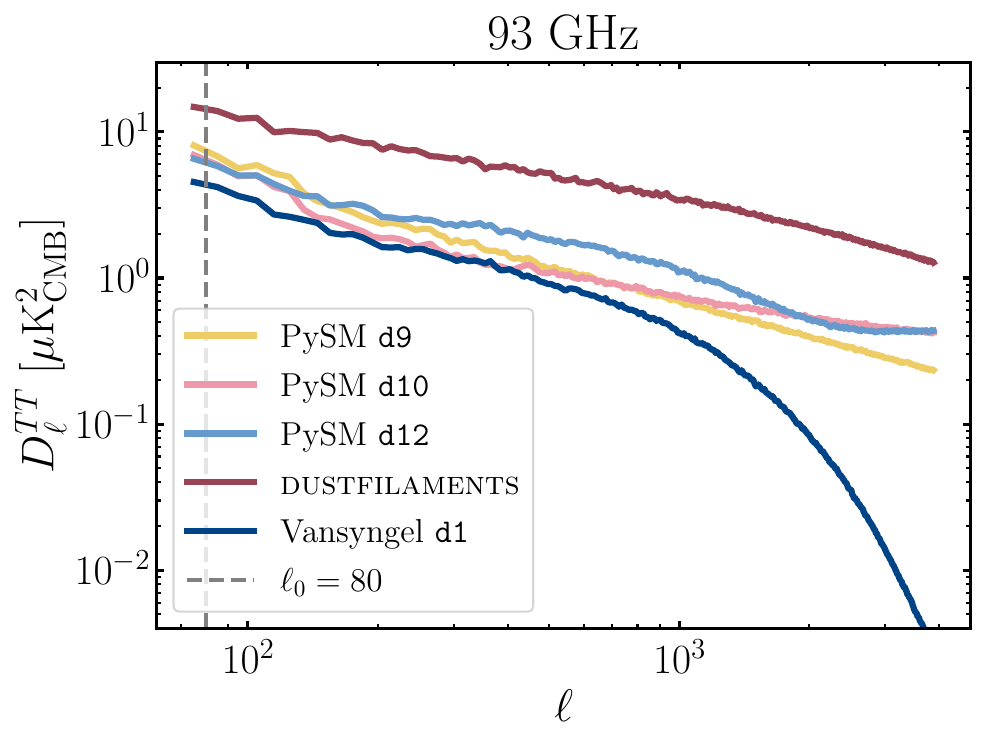}
\caption{\emph{TT} power spectra at \SI{93}{\giga\hertz}, computed on the SO \SI{70}{\percent} mask, for the variety of dust models used in this work. Even though all models are calibrated using \SI{353}{\giga\hertz} \textit{Planck} data at the pivot scale $\ell_0=80$, we see a spread at the level of \SI{0.4}{dex} in power. This is mainly due to differences in both the sky footprint and template used to calibrate the models, and the spectral energy distribution assumed to extrapolate between frequencies (see text for further details). The power in Vansyngel+17 is truncated above $\ell \sim 1000$ due to the presence of a Gaussian beam with $\sigma_{\textrm{FWHM}} = 30^\prime$. Consequently, the constraints derived from this model will be slightly optimistic.}\label{fig:2pt_dusts_GAL060_90GHz} 
\end{figure}

As described earlier, Galactic foregrounds are dominated by dust emission at frequencies above $\nu \gtrsim \SI{70}{\giga\hertz}$ \cite[see e.g.][]{Plack_2016_X}. At low $\nu \lesssim \SI{70}{\giga\hertz}$ frequencies, synchrotron emission overwhelms the CMB signal. As a result, we must consider synchrotron emission models for SO LF channels. We consider PySM \citep{pysm} models \texttt{s4}, \texttt{s5} and \texttt{s7}. These offer varying levels of complexity regarding the modeling of small scales in terms both of amplitude and spectral parameters. They can be combined with PySM \texttt{d9}, \texttt{d10} and \texttt{d12}, respectively. We follow the PySM convention and denote \texttt{d9s4}, \texttt{d10s5} and \texttt{d12s7} as the PySM \texttt{low}, \texttt{medium} and \texttt{high} complexity models. Vansyngel+17 \citep{Vansyngel_2017} also provides a synchrotron realization, produced in a similar procedure as the one used for dust. We denote the sum of those dust and synchrotron simulations as Vansyngel \texttt{d1s1}.\medskip

\subsection{Estimation of foreground biases on lensing reconstruction measurements}

\begin{figure}
\includegraphics[width = 0.9999\columnwidth]{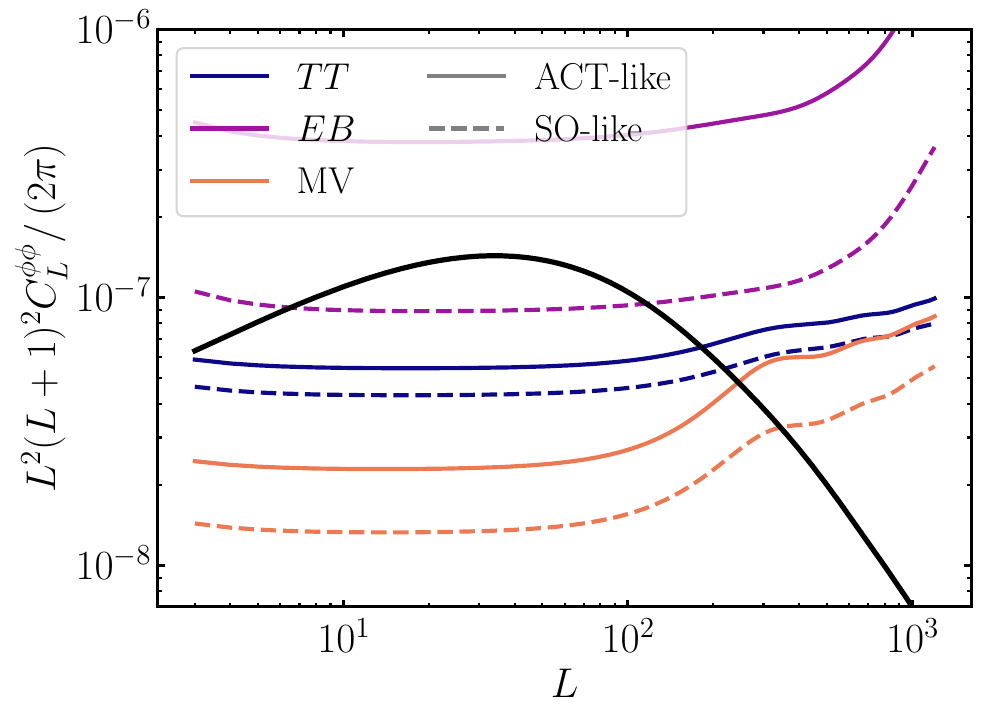}
\caption{Fiducial CMB lensing power spectrum (solid black) compared to the lensing reconstruction noise levels corresponding to different estimators. The $TT$ estimator only offers moderate gain going from an ACT-like experiment (with a white noise level of $\SI{12}{\micro\kelvin}_\mathrm{CMB}$-arcmin) to an SO-like one ($\SI{6}{\micro\kelvin}_\mathrm{CMB}$-arcmin). The vast majority of the improvement comes from estimators making use of polarization channels, such as $EB$. As a result, for a given frequency channel, any increase in foreground bias in temperature-only reconstructions will be primarily driven by the wider area accessed by SO in comparison to ACT.}\label{fig:lensing_noise_levels_SO_ACT} 
\end{figure}

We run the different quadratic estimators on foreground-only maps and measure the autospectrum of the resulting lensing reconstructions. The resulting measurement is $C_L^{\hat{\phi}\hat{\phi},\mathrm{fg}}$. We treat the foreground-only maps in the same way as CMB maps. In particular, we filter the maps (both in temperature and polarization) with the 2-point power defined in equation \eqref{eq:cell}, which does not include the power sourced by the Galactic foreground field. We do this to mimic a real lensing analysis, where Galactic foregrounds are typically not included when constructing the filters. {This does not cause a significant loss of optimality, since the level of dust power predicted by the sky models we use is below the experiment's observed signal and noise power for all relevant multipoles in the analysis. }\medskip

{Because it includes reconstruction noise power, $C_L^{\hat{\phi}\hat{\phi},\mathrm{fg}}$ constitutes an upper limit to the bias that Galactic non-Gaussian foregrounds pose to CMB lensing reconstruction measurements. In particular, current analyses further remove the well-known lensing power spectrum biases, including the $N_L^{(0)}$ and $N_L^{(1)}$ biases;
our results are upper limits because we do not subtract any of those biases in our baseline analysis.} We investigated different avenues to subtract the dominant $N_L^{(0)}$ bias but concluded that we cannot trust our ability to remove it reliably in our estimates (see Appendix \ref{sec:app_patches} for more details).  Hence, in the main text we present upper bounds on the bias produced on the CMB lensing power spectrum amplitude, and, as such, the maximum impact that Galactic foregrounds could produce in our lensing analyses. We follow \cite{Frank_ACT_lensing_2024,MacCrann_2024} and employ the statistic $\Delta A_\mathrm{lens}$ to quantify the bias in the inferred lensing power spectrum amplitude $A_\mathrm{lens}$ due to the presence of Galactic foregrounds:
\begin{equation}\label{eq:deltaAlens_fid}
    \Delta A_\mathrm{lens} = \frac{\sum_L \sigma_L^{-2}  C_L^{\hat{\phi}\hat{\phi},\mathrm{fg}} C_L^{\phi\phi}}{\sum_L \left(C_L^{\phi\phi} / \sigma_L \right)^2},
\end{equation}
{where we assume a fiducial lensing power spectrum signal $C_L^{\phi\phi}$ computed in the default cosmology and a diagonal covariance matrix (with elements $\sigma_L$) representing the error bars of an actual lensing measurement.}\medskip

Since the uncertainty on $A_\mathrm{lens}$ is given by $\left(\sum_L \left(C_L^{\phi\phi}/\sigma_L\right)^2\right)^{-1/2}$, we can also report the bias as a fractional error:
\begin{equation}
    \frac{\Delta A_\mathrm{lens}}{\sigma(A_\mathrm{lens})} = \frac{\sum_L \sigma_L^{-2}  C_L^{\hat{\phi}\hat{\phi},\mathrm{fg}} C_L^{\phi\phi}}{\sqrt{\sum_L \left(C_L^{\phi\phi} / \sigma_L \right)^2}}.
\end{equation}

Since the variance of the estimator is given by the normalization, we will approximate the noise on our lensing bandpowers as
\begin{equation}\label{eq:nl}
    N_L^{\kappa\kappa} = A_L^\phi \left(\frac{L\left(L+1\right)}{2}\right)^2,
\end{equation}
and we will use the Knox formula \citep{Knox_1995, Knox_1997, Allison_2015} to compute the diagonal covariance:
\begin{equation}
    \sigma_L^2 = \frac{2}{ \left(2L+1\right)\Delta Lf_\mathrm{sky}}\left(C_L^{\kappa\kappa} + N_L^{\kappa\kappa}\right)^2,
\end{equation}
where $\Delta L$ is the width of the bandpowers used and $C_L^{\kappa\kappa}$ is our fiducial CMB lensing power spectrum. For ACT, this approach leads to an underestimation of $\sigma(A_\textrm{lens})$, predicting overly optimistic SNRs ($\sim90$) that have not been achieved in practice, due to various reasons including filtering and noise anisotropies. We therefore introduce a common fudge factor of $1.3$ to all our analytical predictions of $\sigma(A_\textrm{lens})$ for ACT to recover the SNR values ($\sim 70$) measured in practice. For SO, this empirical degradation factor is not necessary as $\sigma(A_\textrm{lens})$ is in line with previous forecasts. We use the binning and $L$-range from the latest ACT lensing analysis \citep{Frank_ACT_lensing_2024}: $13$ bins distributed within the range $40<L<1300$ with bin edges $L = [40, 66, 101, 145, 199, 264, 339, 426, 526, 638, 763, 902, 1100,\\ 1300]$. \medskip 

In Figure \ref{fig:lensing_noise_levels_SO_ACT}, we show the fiducial CMB lensing power spectrum (solid black) compared to the lensing reconstruction noise levels corresponding to different estimators. As discussed in $\S$\ref{sec:intro}, we see that the $TT$ estimator only offers moderate gain going from an ACT-like experiment to an SO-like one, {since the CMB temperature is already signal dominated on the relevant scales}. Consequently, for a given frequency channel, any increase in foreground bias in temperature-only reconstructions will be primarily driven to the wider area accessed by SO in comparison to ACT. The vast majority of the improvement comes from estimators making use of polarization channels, such as $EB$. \medskip

\section{Results}\label{sec:results}

\subsection{Temperature-only reconstruction}\label{sec:tonly}

\begin{figure*}
\includegraphics[width = 0.72\textwidth]{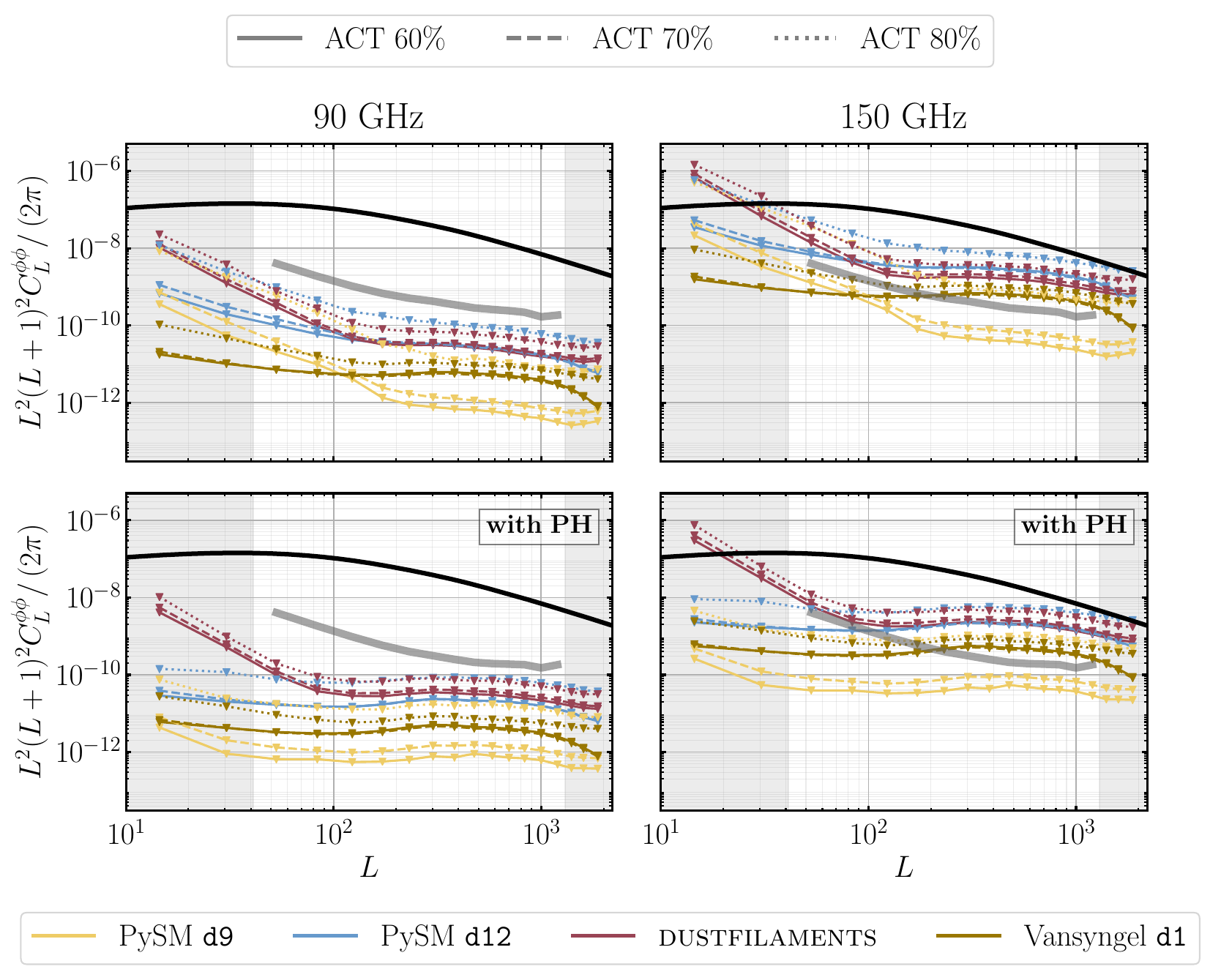}
\includegraphics[width = 0.72\textwidth]{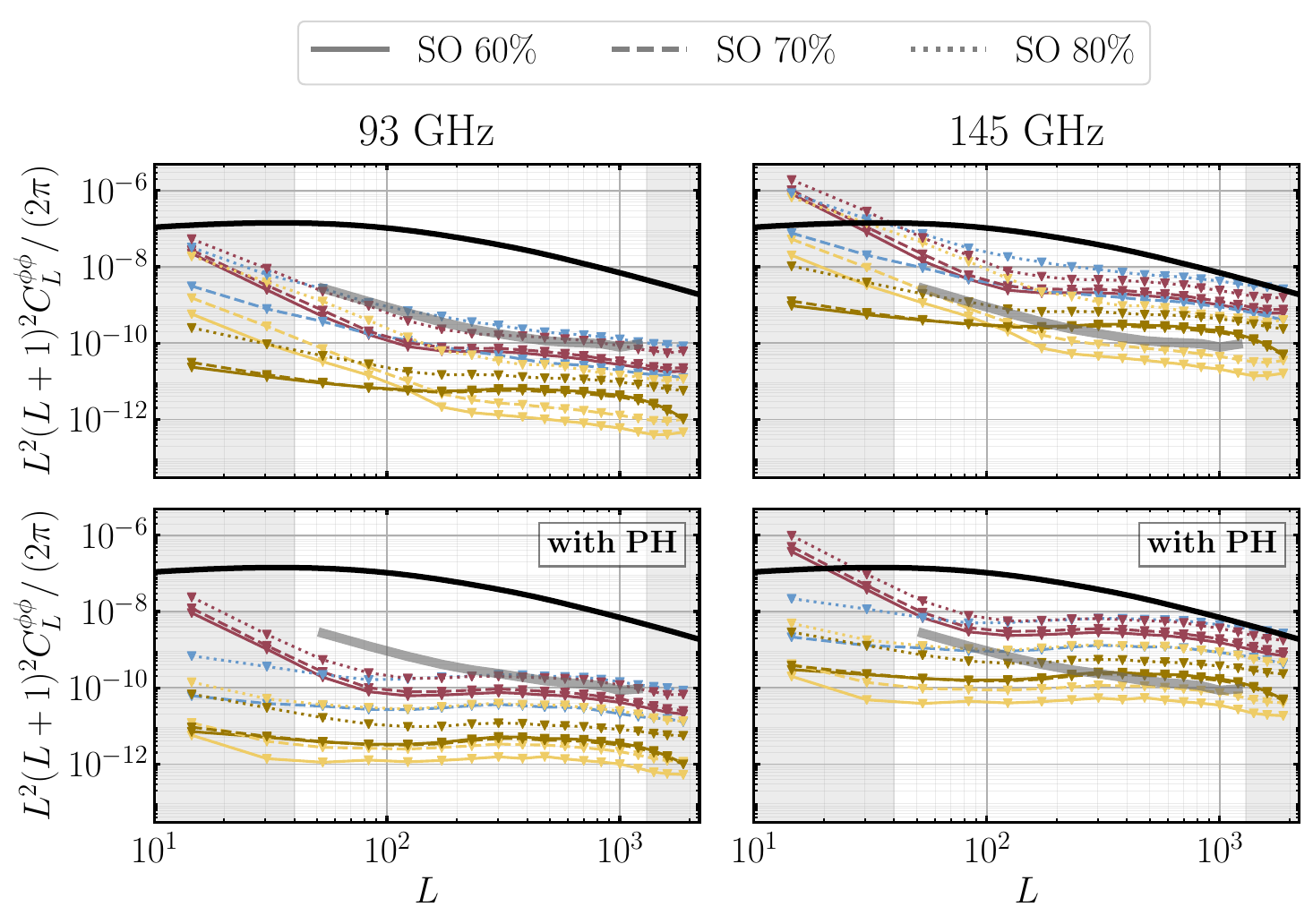}
\caption{Upper limits to the level of Galactic dust contamination on CMB lensing reconstruction measurements in temperature, \iac{for a subset of foreground models studied in this work (see Figure \ref{fig:raw4pt_TT_exp_gals_full} in   Appendix \ref{ap:full_versions} for the full version of this figure)}. We use the scales $600<\ell_T<3000$ for the reconstruction. For reference, we also show the fiducial CMB lensing powers spectrum (black line), and mark as a thick gray line the level of foreground that would produce a 0.5$\sigma$ shift on the inferred CMB lensing power spectrum amplitude, which varies depending on the experimental configuration and the estimator used. The white regions correspond to the scales usually used to derive cosmological parameters, namely $40 < L < 1300$. The SED of dust is such that biases in the lensing reconstruction are almost $2$ orders of magnitude larger in the $\SI{150}{\giga\hertz}$ channel (right column) than in the $\SI{90}{\giga\hertz}$ one (left column). Biases also increase as a function of sky fraction, denoted with different line styles \iac{(for clarity, we have also omitted PySM \texttt{d12} results on the SO 60\% mask in this figure).} The top (bottom) four panels correspond to results with an experimental setup similar to ACT (SO). The panels labeled as ``with PH'' include profile hardening (tSZ-optimized) as a mitigation strategy, although this does not decrease of the level of contamination in the scales of interest significantly.}\label{fig:raw4pt_TT_exp_gals} 
\end{figure*}

\begin{figure*}
    \centering
     \includegraphics[width=0.4955\linewidth]{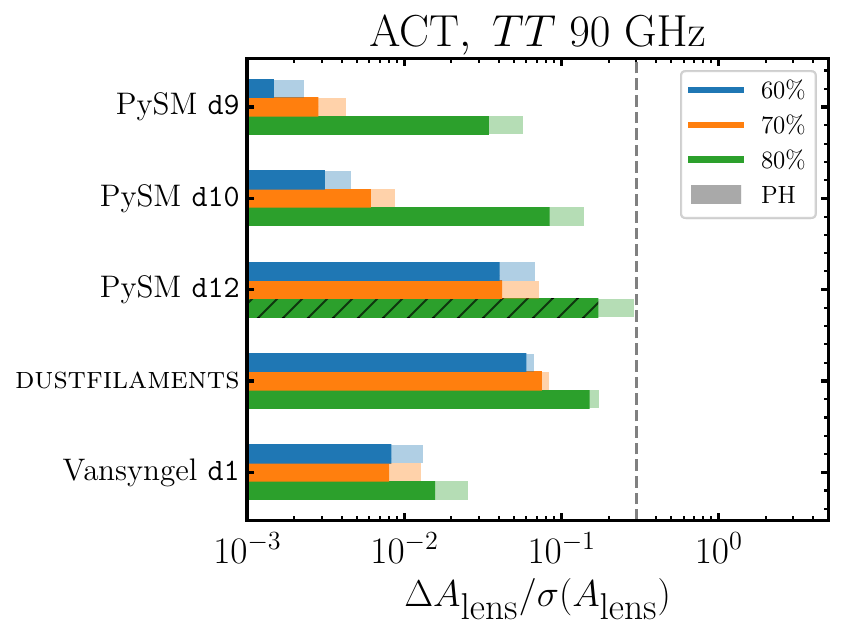}
    \includegraphics[width=0.4955\linewidth]{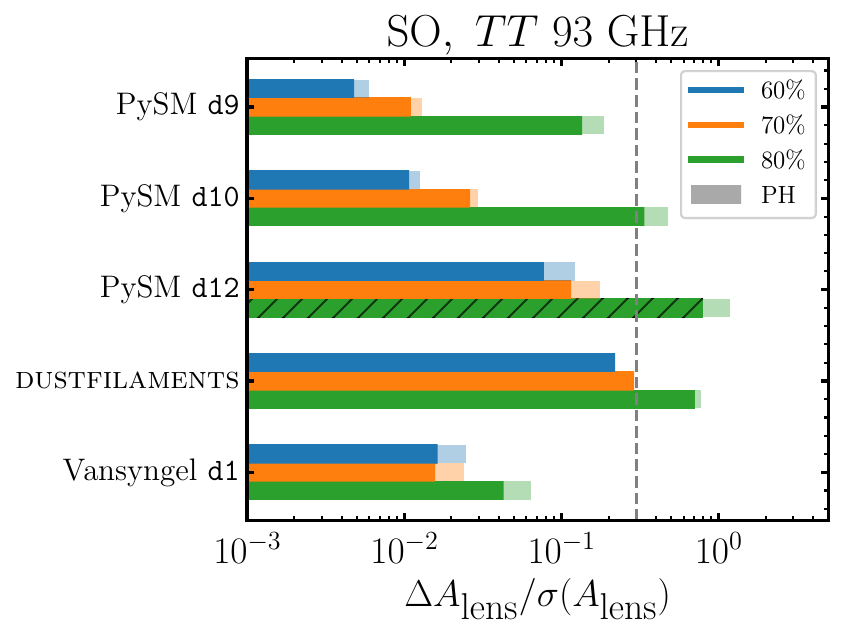}
    \caption{Summary of upper limits to the bias of the CMB lensing power spectrum amplitude temperature-only reconstruction induced by the different dust models \iac{on ACT (left panel) and SO (right panel)}. We show as different colors the results for different sky fractions (60\%: blue, 70\%: orange, 80\%: green). The hatched bars correspond to results obtained on the same setting but with profile-hardening (PH) added as a mitigation strategy. This helps in all models except on \textsc{dustfilaments}. For reference, with profile hardening, the uncertainty on $A_\textrm{lens}$ on the 70\% masks is $0.019$ (ACT) and $0.011$ (SO). We find that the shift induced on $A_\textrm{lens}$ remains well below $0.3\sigma$ (vertical dashed line) for ACT, and also for SO when excluding the 80\% mask. We show results at $\SI{90}{\giga\hertz}$. At $\SI{150}{\giga\hertz}$, the biases are in all cases almost a hundred times larger.}
    \label{fig:alenstable_90GHz}
\end{figure*}

In Figure \ref{fig:raw4pt_TT_exp_gals}, we show the results of running the $TT$ quadratic estimator on different non-Gaussian dust foreground models, and as a function of sky fraction. The top panel corresponds to ACT-like footprints and noise levels, whereas the bottom panel refers to SO-like ones. We also show the fiducial CMB lensing power spectrum for reference as a solid black line. These curves present an upper bound on dust contamination. We have not subtracted any known biases, and hence their shape at low $L$ is contaminated by the mean-field bias produced by the mask. On scales much smaller than the ones shown in these Figures, we see the characteristic shape produced by the $N_L^{(0)}$ bias. An explanation of these is presented in Figure \ref{fig:explain_biases_wgauss} (Appendix \ref{sec:app_QE}). It is not straightforward to estimate and subtract those biases because, on the one hand, the two-point power spectrum used to compute the analytical $N_L^{(0)}$ bias is not representative of the local two-point power spectrum across the map. On the other hand, we do not have enough realizations of the dust field for the different models to estimate the mean-field. In Appendix \ref{sec:app_patches}, we explore a possible strategy based on a patch-by-patch reconstruction, which has previously been proposed in \citep{Namikawa_2013, Challinor_2018}. \iac{This approach had previously been used in \citep{Planck13_lensing}, following \citep{Plaszczynski_2012}, as a validation method of their main reconstruction pipeline}.\medskip

The white regions in Figure \ref{fig:raw4pt_TT_exp_gals} correspond to the scales used in \citep{Frank_ACT_lensing_2024} to derive cosmological parameters, namely $40 < L < 1300$, and hence the scales on which we compute the upper limit to the bias $\Delta A_{\textrm{lens}}$ on the lensing power spectrum amplitude $A_{\textrm{lens}}$. We report these in Figure \ref{fig:alenstable_90GHz}, in terms of the measurement uncertainty on $A_\textrm{lens}$. On the 70\% masks, $\sigma(A_\textrm{lens})$ is $0.016$ (ACT) and $0.009$ (SO). In these Figures \ref{fig:raw4pt_TT_exp_gals}--\ref{fig:alenstable_90GHz}, we focus on two different mid-frequency channels where the CMB is brightest. Since the dust intensity at \SI{150}{\giga\hertz} is $\sim 3$ times brighter than in the \SI{90}{\giga\hertz} channel (see $\S$\ref{sec:skymodel}), we see that the biases in the lensing reconstruction are almost $2$ orders of magnitude larger in the $\SI{150}{\giga\hertz}$ channel than in the $\SI{90}{\giga\hertz}$ one.\footnote{{Fortunately for ACT lensing, the $\SI{90}{\giga\hertz}$ frequency channel has more statistical weight than the $\SI{150}{\giga\hertz}$ one, as we discuss in \S\ref{sec:ilc}.}} Finally, we perform lensing reconstruction on different footprints. As expected, the biases increase with sky fraction since we are accessing regions closer to the Galactic plane, where Galactic foreground emission dominates. At $\SI{90}{\giga\hertz}$, we find that the shift induced on $A_\textrm{lens}$ remains below $0.3\sigma$ for ACT ({limit marked as a vertical dashed gray line in Figure \ref{fig:alenstable_90GHz}}). It is well below the $0.5\sigma$ level for SO ({limit marked as a thick gray line in Figure \ref{fig:raw4pt_TT_exp_gals}}), except for when using the 80\% mask {for which biases become non-negligible}. As a result (unless there are further improvements to our conservative upper limits in future work), we advise against using \textit{Planck} \SI{80}{\percent} dust mask for CMB lensing reconstruction analyses\footnote{\iac{We also note that \citep{Planck13_lensing} found discrepancies at low $L$ when using $f_\textrm{sky}=0.8$, which was suspected to be result of a residual dust contamination.}}.\medskip

\subsubsection{Mitigation strategies: profile hardening}

A common mitigation strategy of extragalactic foregrounds in CMB lensing analyses is the use of bias-hardened quadratic estimators. This is a geometric method that modifies the form of the standard QE such that, at leading order, it has zero response to the statistical anisotropy caused by these foregrounds \citep{Namikawa_2013, Osborne_2014, Sailer_2020, Sailer_2023}. In particular, \cite{Frank_ACT_lensing_2024, MacCrann_2024} use a tSZ-like profile to bias-harden against tSZ clusters, although the method performs well in the presence of other foregrounds with different profiles, such as the CIB \citep{Sailer_2023}. We test whether the tSZ-profile-hardened ``source'' ($s$) estimator also successfully mitigates non-Gaussian dust foregrounds. {Since the Galactic dust distribution contains both source-like clumps and a large-scale mask-like modulation in its amplitude, we might expect hardening approaches, which reduce the impact of both point sources and mask effects, to reduce the level of dust bias.}\medskip

In the presence of this foreground, equation \eqref{eq:TT} becomes:
\begin{equation}
   \langle T_\mathbf{l} T_{\mathbf{L}-\mathbf{l}} \rangle= f^\phi_{\mathbf{l}, \mathbf{L}} \phi_\mathbf{L} + f^s_{\mathbf{l}, \mathbf{L}} s_\mathbf{L},
\end{equation}
where the average is taken at fixed lensing and source field. The bias-hardened estimator for $\phi_\mathbf{L}$ in the presence of {the source field} $s_\mathbf{L}$ is \citep{Namikawa_2013, Osborne_2014, Sailer_2020}:
\begin{equation}
    \phi_\mathbf{L}^{\textrm{BH},s} = \frac{1}{1 - \mathcal{R}_\mathbf{L}^{\phi s} \mathcal{R}_\mathbf{L}^{s\phi}} \left( \hat{\phi}_\mathbf{L} - \mathcal{R}_\mathbf{L}^{\phi s} \hat{s}_\mathbf{L}\right),
\end{equation}
where $\hat{s}_\mathbf{L}$ is an estimator for the source field, constructed following the same procedure as that for the lensing field $\phi_\mathbf{L}$, and $\mathcal{R}_\mathbf{L}^{\phi s}$ is the response function \citep{Namikawa_2013}:
\begin{equation}\label{eq:response}
    R_\mathbf{L}^{a,b} \sim A_\mathbf{L}^a \int \mathrm{d}^2\mathbf{l} f^{a}(\mathbf{l}, \mathbf{L}) f^b(\mathbf{l}, \mathbf{L}),
\end{equation}
which we define to absorb the normalization of the estimator. There is a moderate noise cost for bias hardening (BH), where the noise in the reconstruction (c.f. equation \ref{eq:nl}) now takes the form \citep{Namikawa_2013, Osborne_2014, Sailer_2020,MacCrann_2024}:
\begin{equation}\label{eq:nlkkbh}
    N_L^{\kappa\kappa, \textrm{BH}} = N_L^{\kappa\kappa}\left(1 - N_L^{\kappa\kappa} N_L^{ss} \mathcal{R}_L^2\right)^{-1}, 
\end{equation}
where $\mathcal{R}_L$ is the response of the estimator to the presence of the foreground we harden against and $N_L^{ss}$ is the noise of the source estimator. With profile hardening, our estimate of $\sigma(A_\textrm{lens})$ increases by \SI{20}{\percent}: $0.019$ (ACT) and $0.011$ (SO).\medskip


The results are shown in the panels labeled as ``with PH'' in Figure \ref{fig:raw4pt_TT_exp_gals}. We also quote the bias in the lensing amplitude as hatched bars in Figure \ref{fig:alenstable_90GHz}. We find that profile-hardening helps reduce the amplitude of the bias on the PySM and Vansyngel+17 models, although the improvement is not very large. This result is somewhat surprising since the bias produced by dust ``blobs'' should be absorbed by profile hardening, as noted earlier. We find the improvement to be mainly driven by a suppression of the power spectrum amplitude at low $L$, usually enhanced by the presence of a mask. We can see this explicitly by computing the expected value of the bias-hardening source estimator:
\begin{equation}\label{eq:mf}
    \langle\phi_\mathbf{L}^{\textrm{BH},s}\rangle = \frac{1}{1 - \mathcal{R}_\mathbf{L}^{\phi s} \mathcal{R}_\mathbf{L}^{s\phi}} \left( \langle\hat{\phi}_\mathbf{L} \rangle - \mathcal{R}_\mathbf{L}^{\phi s} \langle\hat{s}_\mathbf{L}\rangle\right),
\end{equation}
in the presence of both foreground $s$ and a mask $w(\nv)$, that is:
\begin{equation}\label{eq:tt_phisf}
      \langle T_\mathbf{l} T_{\mathbf{L}-\mathbf{l}} \rangle= f^\phi_{\mathbf{l}, \mathbf{L}} \phi_\mathbf{L} + f^s_{\mathbf{l}, \mathbf{L}} s_\mathbf{L} + f^M_{\mathbf{l}, \mathbf{L}} M_\mathbf{L},
\end{equation}
where we have introduced the complementary function of the mask in Fourier space, $M_\ell = \delta_\ell - W_\ell$  \citep{Namikawa_2013}. We introduce \eqref{eq:tt_phisf} in \eqref{eq:phihat} and substitute in \eqref{eq:mf}. Making use of \eqref{eq:response}, we find:
\begin{equation}
    \langle\phi_\mathbf{L}^{\textrm{BH},s}\rangle = \phi_\mathbf{L} + \frac{\mathcal{R}^{\phi M}_\mathbf{L} - \mathcal{R}_\mathbf{L}^{\phi s} \mathcal{R}_\mathbf{L}^{sM}}{1 - \mathcal{R}_\mathbf{L}^{\phi s} \mathcal{R}_\mathbf{L}^{s\phi}} M_\mathbf{L}.
\end{equation}
That is, the prefactor quantifying the contribution to the bias due to the presence of a mask (originally $\mathcal{R}^{\phi M}_\mathbf{L}$) has now been reduced. This reduction is proportional to how similarly the two fields respond to the presence of $\phi$. In the limit $s\sim M$, $\mathcal{R}^{s M}_\mathbf{L} = 1$, and the contribution from the mask is completely suppressed. 
We compute explicitly the ratio of this new prefactor obtained when we bias-harden against point sources with respect to the original response $\mathcal{R}^{\phi M}_\mathbf{L}$. The reduction is typically 20\%, with values ranging from 30\% at $L\sim10$ to 10\% at $L \sim 1000$.\medskip

This idea is now linked to the fact that both PySM and Vansyngel+17 models rely on modulating small scales by a large-scale field $t(\nv)$ to generate non-Gaussianities. This operation can straightforwardly be absorbed by the presence of a mask $\tilde{w}(\nv) = w(\nv)t(\nv)$. Therefore, using profile-hardening will also mitigate against biases arising from this new mask ($M_\ell$ now becomes $\tilde{M}_\ell = \delta_\ell - \tilde{W}_\ell$ in the discussion above). {\textsc{dustfilaments} do not synthesize small scales following the modulation approach. We can speculate that since \textsc{dusfilaments} neither looks like a mask modulation nor looks like individual sources, instead appearing as very elongated structures (as can be seen in Figure \ref{fig:cutouts}), profile hardening has minimal impact.}\medskip

In order to decrease the bias at all multipoles, new profiles based on the foreground's bispectrum and power spectrum \citep{Sailer_2020} are needed. For example, these could be chosen to reflect the typical size of filaments or dust clumps. We leave this line of research to future work.\medskip

\begin{figure*}
\includegraphics[width = 0.85\textwidth]{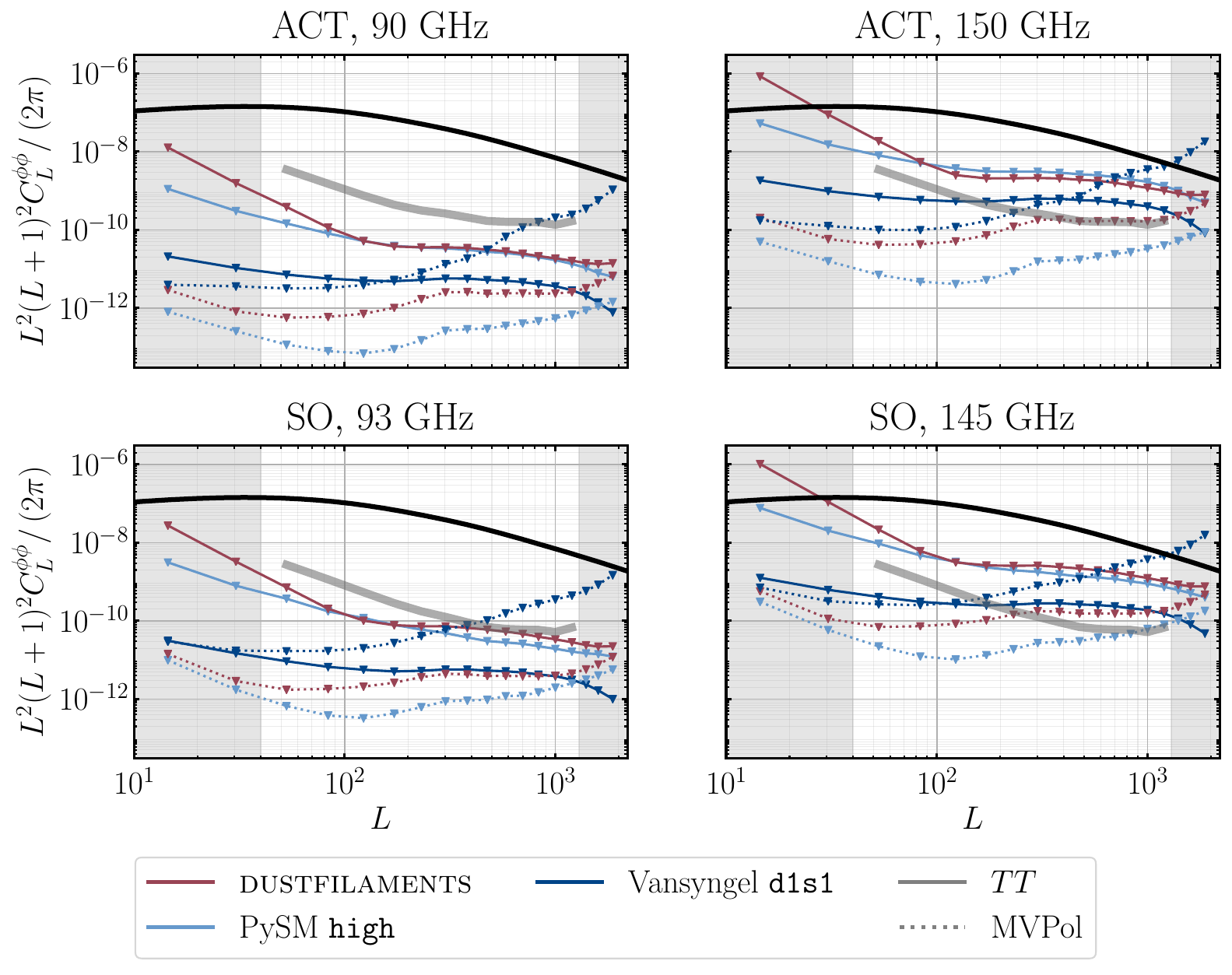}
\caption{Upper limit to the level of Galactic foreground contamination on CMB lensing reconstruction measurements \iac{for the temperature and polarization-only quadratic estimators} (marked with different linestyles). The top panels show results for ACT on the ACT \SI{70}{\percent} footprint  ($f_\textrm{sky} = 0.31$), and the bottom panels correspond to SO on the SO \SI{70}{\percent} footprint ($f_\textrm{sky} = 0.42$). The plots on the left (right) correspond to the \SI{90}{\giga\hertz} (\SI{150}{\giga\hertz}) channels. Different colors correspond to different foreground models. The reconstruction is performed with multipoles $600<\ell<3000$, and we shade the $L$ regions excluded on cosmological analyses. For reference, we mark as a thick gray line the level of foreground that would produce a 0.5$\sigma$ shift on the inferred CMB lensing power spectrum amplitude for the different configurations with the minimum-variance estimator. On PySM and \textsc{dustfilaments} maps, estimators that make use of polarization measurements are less biased, whereas Vansyngel \texttt{d1s1} model presents a higher bias in polarization than in temperature. At $\SI{150}{\giga\hertz}$, the biases show similar trends than at $\SI{90}{\giga\hertz}$ but are almost a hundred times larger, as was discussed in \S\ref{sec:tonly}.}
\label{fig:ACT_ests} 
\end{figure*}

\subsection{Galactic foreground biases in both intensity and polarization} 

As CMB experiments reach higher sensitivities, polarization measurements will contribute to the majority of SNR in lensing analyses \citep{Pan_SPT3G_2023_lensing, Ge_SPT_2024} and quadratic estimators will become suboptimal \citep{Hirata_2003, Carron_MAP_2017, Legrand_2022, Legrand_2023}. The improvements from optimal methods applied to experiments such as ACT or SO are very modest. We do expect, however, polarization measurements to contribute significantly to the reconstruction SNR, especially for SO. As a result, we now explore the impact of non-Gaussian Galactic foregrounds in both intensity and polarization within the framework of quadratic estimators. We now include synchrotron within our foreground modeling (PySM \texttt{low}, \texttt{medium} and \texttt{high} complexity models, as discussed in $\S$\ref{sec:skymodel}), even though its contribution on frequencies above \SI{70}{\giga\hertz} is negligible. We further check that synchrotron produces no shift in the CMB lensing power spectrum bias on those channels.\medskip

\begin{figure*}
\includegraphics[width = 0.95\textwidth]{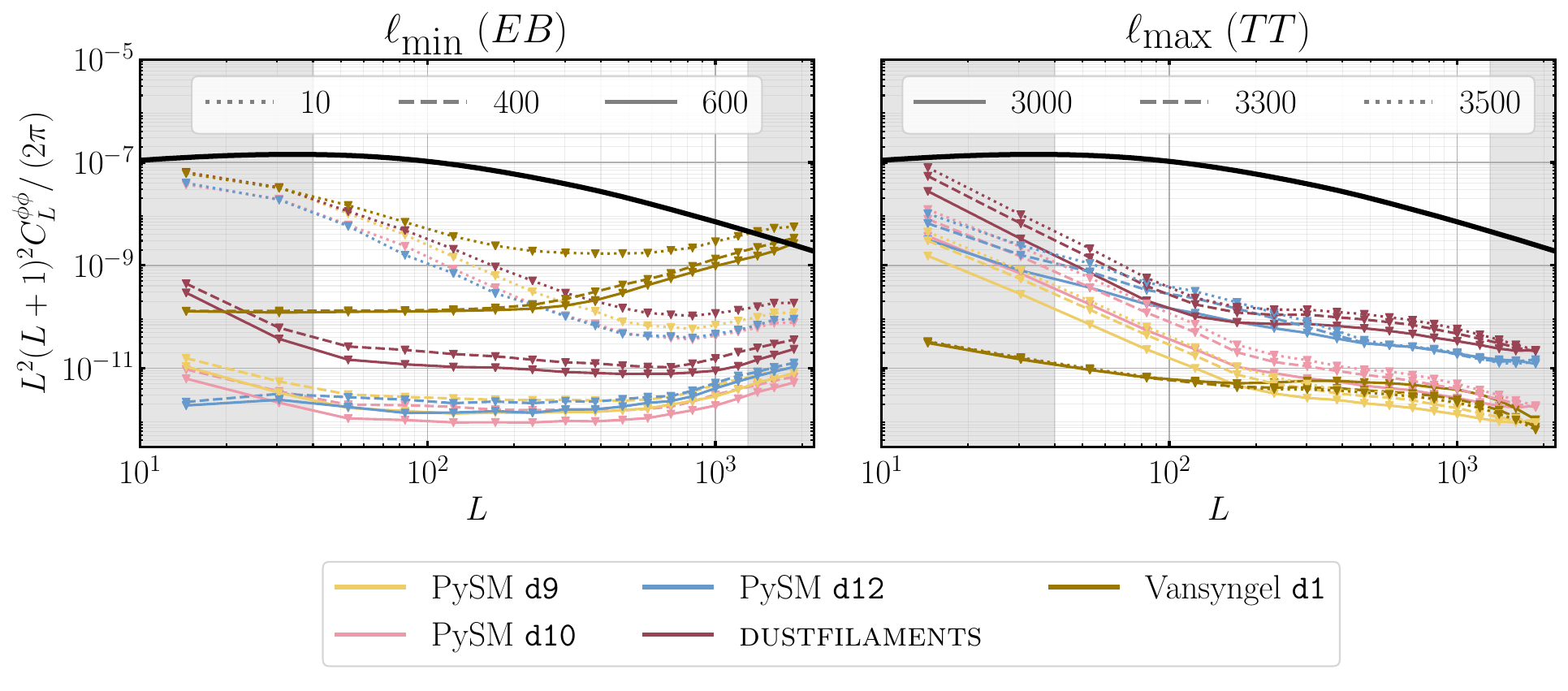}
\caption{Upper limit on the raw lensing power spectrum due to non-Gaussian dust foregrounds in an experiment such as SO (\SI{70}{\percent} mask, $f_\textrm{sky}=0.42$) at \SI{93}{\giga\hertz}, as a function of the multipole range used in the reconstruction, for $TT$ and $EB$ estimators. \textit{Left}: As $\ell_{\textrm{min}}$ is varied (fixed $\ell_{\textrm{max}} = 3000$), the $EB$ estimator presents significant shifts in the amplitude of the bias because it has important contributions from low-$\ell$ modes. \textit{Right}: Increasing $\ell_{\textrm{max}}$ (while keeping $\ell_{\textrm{min}} = 600$ fixed) in the temperature-only estimator also leads to an increase in the bias, although the shifts are small.}\label{fig:raw4pt_TT_SO_lminlmax} 
\end{figure*}

In Figure \ref{fig:ACT_ests}, we show upper limits to the CMB lensing power spectrum bias \iac{for the temperature and polarization-only quadratic estimators}. The top panel shows results for ACT channels on the ACT \SI{70}{\percent} footprint  ($f_\textrm{sky} = 0.31$), and the bottom panel corresponds to the SO mid-frequency channels on the SO \SI{70}{\percent} footprint ($f_\textrm{sky} = 0.43$). We only include the most complex PySM model (PySM \texttt{high}), together with \textsc{dustfilaments} and Vansyngel \texttt{d1s1} for clarity. Firstly, we find that on PySM and \textsc{dustfilaments} maps, estimators that make use of polarization measurements are less biased. This is in agreement with previous works \citep{Challinor_2018}, and it is not a result of the lack of $EB$ correlation present on these models. Any non-zero $EB$ signal cannot produce a mean-field bias in the $EB$ estimator. However, it could generate some non-zero bias at the trispectrum level, as pointed out by \citep{Beck_2020}. We show in Appendix \ref{sec:app_eb} that the level of $EB$ correlations compatible with existing data are too small to have any detectable contribution to the trispectrum.\medskip

Finally, we also find that Vansyngel+17 models present a higher bias in polarization than in temperature. We believe this is due to the loss of power above $\ell\sim 1000$ present in these simulations, which we explore further in the next section (\S\ref{sec:multipole}). In a nutshell, the temperature-only estimator receives most of its contribution from high multipoles \citep{Schmittfull_2013}, but there is no signal on those small scales on this model. \medskip 

Previous papers have found that the biases present in polarization due to non-Gaussian dust foregrounds can be mitigated well below the experiment's requirements after including the dust power spectrum in the filters for the CMB fields (equation \ref{eq:cell}). This strategy did not work for the temperature-only estimator since the CMB component dominates the temperature power spectrum CMB \citep{Challinor_2018, Beck_2020}. We are in the same position here. The level of dust predicted by the sky models we use is below the experiment's observed signal (including noise) for all relevant multipoles in the analysis, so including it does not serve as a mitigation strategy.\medskip

\subsubsection{Mitigation strategies: choice of multipole range used in the reconstruction}\label{sec:multipole}\medskip

We follow \citep{Fantaye_2012, Beck_2020} and show the sensitivity of the bias to changing the multipole range used in the lensing reconstruction. We explore changes in both $\ell_{\mathrm{min}}$ and $\ell_{\mathrm{max}}$ with respect to the default analysis range, $600<\ell<3000$ \citep{Frank_ACT_lensing_2024}. In particular, Galactic foreground biases are known to worsen when decreasing the minimum multipole. \medskip

We show in Figure \ref{fig:raw4pt_TT_SO_lminlmax} the upper limit of dust foreground biases in the SO-like experiment (\SI{70}{\percent} mask) configuration. The left panel shows the sensitivity to lowering $\ell_{\textrm{min}}$ in polarization, while the right panel corresponds to changes in  $\ell_{\textrm{max}}$ in a temperature-only reconstruction. We report changes in $A_\textrm{lens}$ always with respect to the baseline $\ell_{\textrm{min}}=600$, $\ell_{\textrm{max}}=3000$. In polarization, the bias on $A_\textrm{lens}$ increases from  $0.8\sigma$ at $\ell_{\textrm{min}}=600$ up to $6.8\sigma$ for $\ell_{\textrm{min}}=10$ within the Vansyngel \texttt{d1} dust model.\footnote{Nevertheless, in this case the bias can be reduced to a $0.8\sigma$ bias by simply including the foreground power spectra in the filters. When multipoles below $\ell\sim 100$ in polarization are considered, the dust power spectrum is greater than the observed CMB signal (including noise), so it would be suboptimal not to include them in the filters.} In temperature, increasing $\ell_{\textrm{max}}$ increases the amplitude of the bias, although the shift is minimal (from $0.3\sigma$ to $0.8\sigma$, corresponding to the \textsc{dustfilaments} model as one goes from $\ell_{\textrm{max}}=3000$ to $\ell_{\textrm{max}}=3500$). We see no effect of changing the minimum multipole used in the reconstruction for the $TT$ estimator, or the maximum multipole for the $EB$ estimator, so these are not shown.\medskip

These results can be explained as follows: the contribution of CMB multipoles $\ell \lesssim 700$ to the signal-to-noise ratio of the lensing reconstruction in temperature is very small \citep{Schmittfull_2013}, so we only see appreciable changes when we vary the maximum multipole instead. On the smallest scales, however, Galactic foregrounds become subdominant and produce a minimal change if included. The $EB$ estimator, on the other hand, has significant contributions from lower $\ell$ \citep{Pearson_2014}, the scales at which Galactic foregrounds become more important. \medskip 

Similar behavior was found in \citep{Fantaye_2012}, who suggest low-$\ell$ filtering as a useful robustness check for the effect of diffuse foreground biases. As a general guidance, CMB lensing analyses should always ensure the consistency of different scale cuts.\medskip 

In this section, we have stressed how high-$\ell$ multipoles contribute the most to the signal-to-noise in temperature-only reconstructions \citep{Schmittfull_2013}. These scales, however, are not present in the Vansyngel+17 model due to the presence of a $30^\prime$ beam. If this was not the case, we would expect this model to present a larger bias in temperature than in polarization, as it is the case for the other sky models explored in this paper. This is apparent in Figure \ref{fig:raw4pt_TT_SO_lminlmax}: both solid lines in the two plots correspond to the same multipole range used in the reconstruction ($600<\ell<3000$), and we see that $TT$ presents a larger bias in temperature than in polarization in all models except the Vansyngel+17 one.  As a result, the constraints derived with the Vansyngel+17 model \citep{Vansyngel_2017} for both temperature and minimum variance estimators are slightly optimistic. \medskip

\subsection{Biases on coadded frequency channels}\label{sec:ilc}\medskip

\begin{figure}
\includegraphics[width = 0.9\columnwidth]{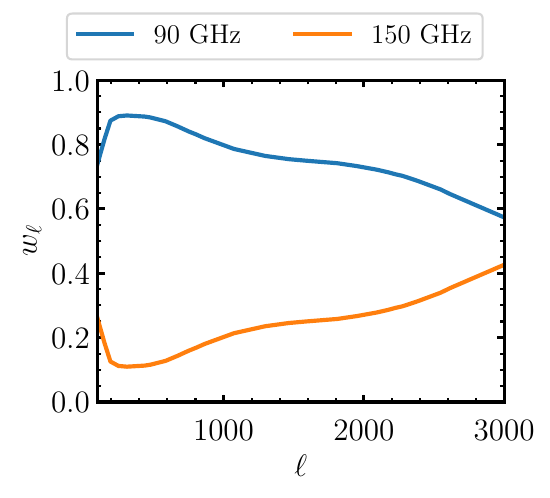}
\caption{Inverse-variance weights used to combine ACT \SI{90}{} and \SI{150}{\giga\hertz} channels in temperature. They are derived from the actual ACT noise power spectra \citep{Atkins_2023}, and are normalized to unity at every multipole $\ell$. Similar weights are computed to combine the polarization ($E$, $B$) components. The \SI{90}{\giga\hertz} channel provides the largest contribution to the coadded map.}\label{fig:weights_night} 
\end{figure}

\begin{figure}
\includegraphics[width = 0.9999\columnwidth]{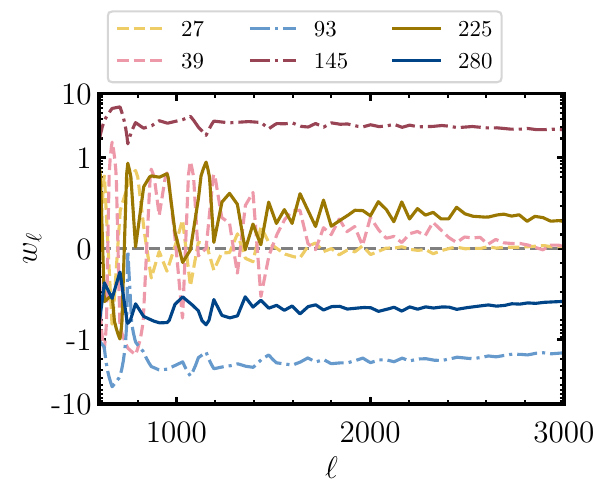}
\caption{HILC weights used to combine all six SO frequency channels in temperature (labeled in \SI{}{\giga\hertz} units). They are computed directly from the data covariance, encoding all auto- and cross-spectra (including the lensed $TT$ CMB spectra, thermal and kinetic SZ effects, CIB and radio point source emission \citep{SO_2019}). For clarity, we smooth the resulting weights before plotting them. Similar weights are computed to combine the polarization ($E$, $B$) components. The mid-frequency channels provide the dominant contribution to the HILC map.}\label{fig:sohilc_weights} 
\end{figure}

In practice, CMB lensing measurements are not performed on individual frequency maps. The first step is often to combine the data gathered by the experiment at the different frequencies into a single CMB map. This has several advantages, including foreground mitigation. We now discuss biases measured on coadded frequency channels. This method is similar to the foreground-cleaning approaches discussed in \citep{Fantaye_2012, Challinor_2018, Beck_2020}, which were found to significantly help reduce Galactic foreground biases.\medskip

\begin{figure*}
\includegraphics[width = 0.95\textwidth]{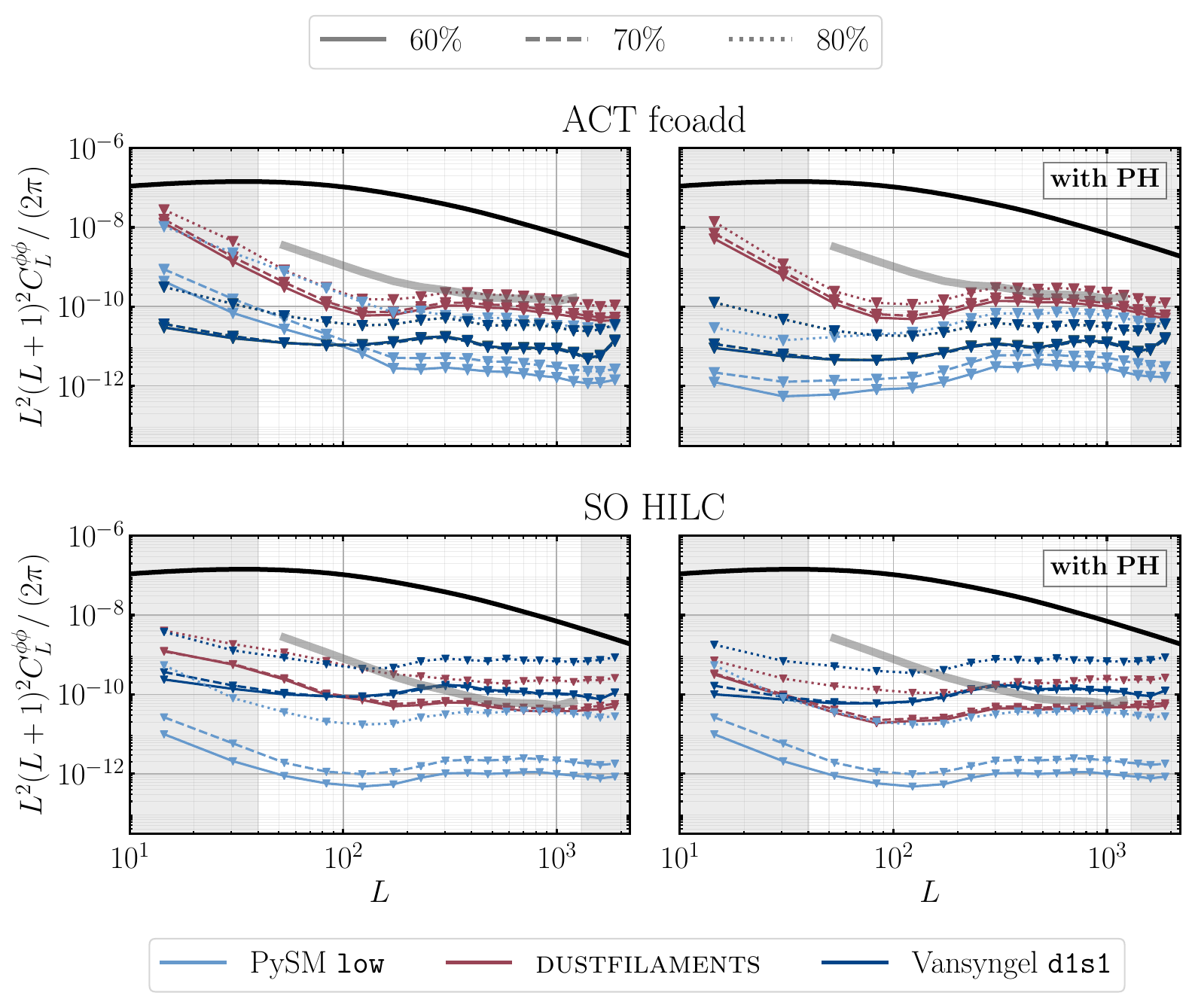}
\caption{Upper limits on the amplitude of the non-Gaussian Galactic foreground bias using the minimum-variance quadratic estimator within ACT and SO, for \iac{a subset of the foreground models studied in this work and as a function of sky fraction (see Figure \ref{fig:raw4pt_dusts_GALS_HILC} in Appendix \ref{ap:full_versions} for the full version of this figure)}. We include profile hardening (tSZ-optimized) in the panels marked as ``with PH''. For ACT, we use inverse-variance weighting coaddition of frequencies. For SO, we use the HILC method to coadd all six frequency channels. After coadding, and within the \SI{70}{\percent} mask region, these upper limits on the Galactic foregrounds biases are at the sub-percent level for the ACT experiment. We quantify the bias on $A_\textrm{lens}$ on the left panel of Figure \ref{fig:alenstable_ilc}. For reference, we mark as a thick gray line the level of foreground that would produce a 0.5$\sigma$ shift on the inferred CMB lensing power spectrum amplitude for the different configurations.}\label{fig:raw4pt_dusts_GALS_HILC} 
\end{figure*}

\begin{figure*}
    \includegraphics[width=0.4955\linewidth]{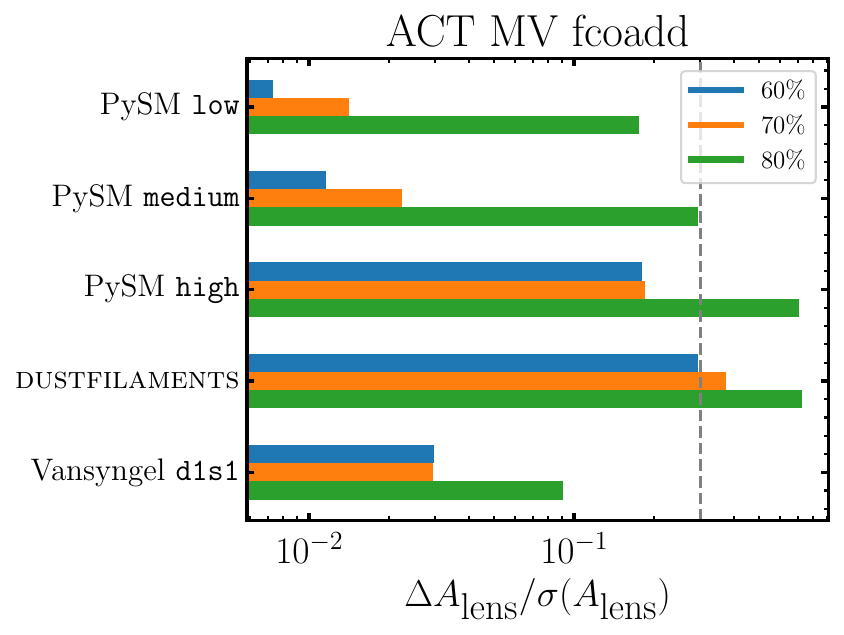}
    \includegraphics[width=0.4955\linewidth]{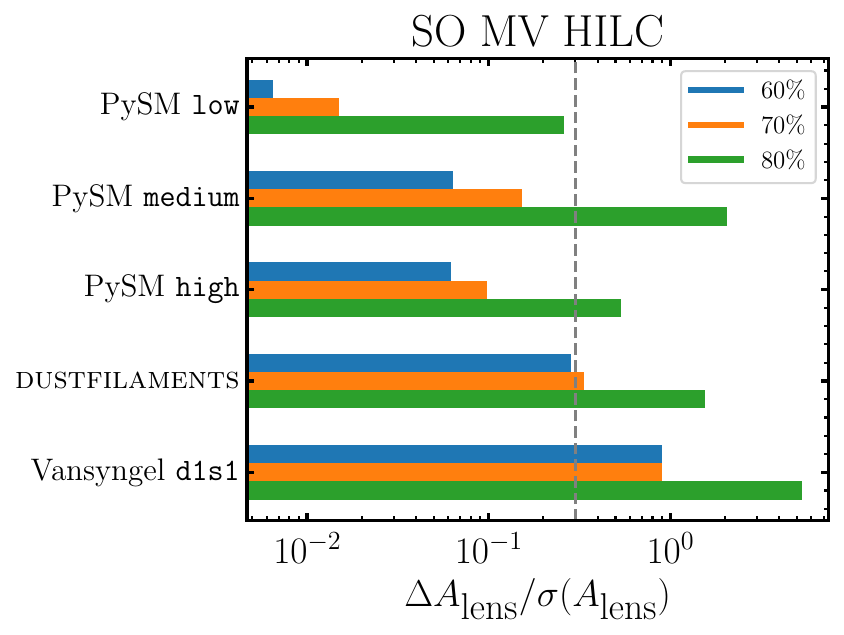}
    \includegraphics[width=0.4955\linewidth]{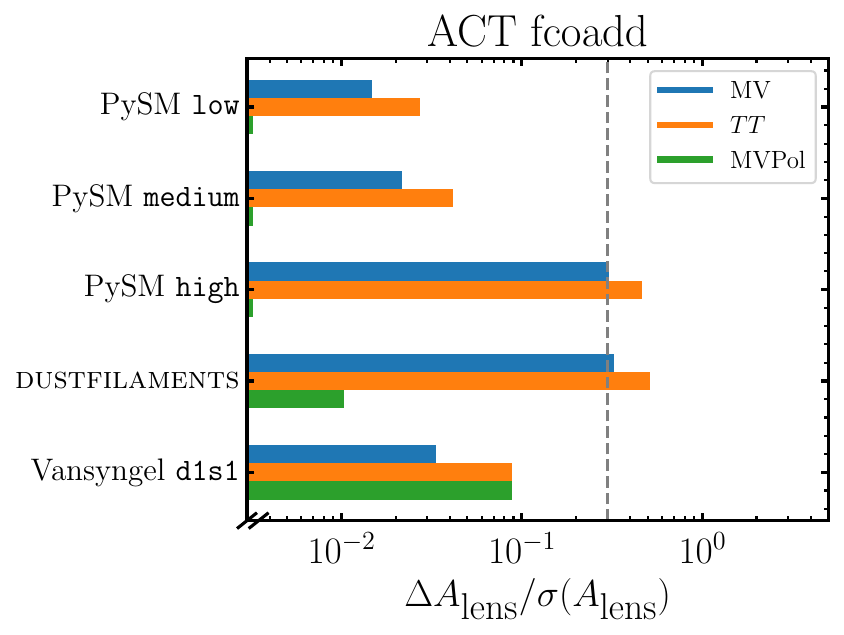}
    \includegraphics[width=0.4955\linewidth]{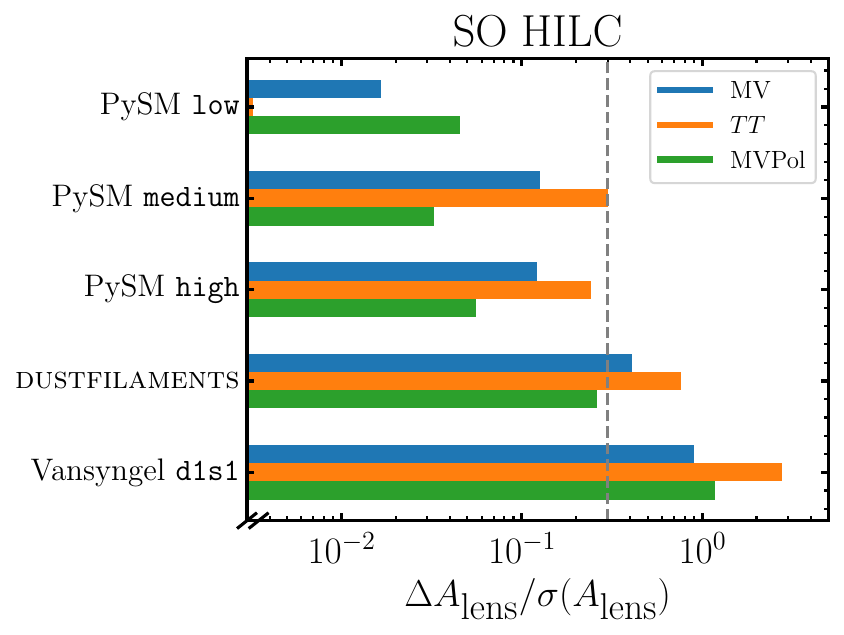}
    \caption{Similar to Figure \ref{fig:alenstable_90GHz}, summary of upper limits to the bias of the CMB lensing power spectrum amplitude reconstruction induced by the different dust models, \iac{for both ACT (left column) and SO (right column) experiments}, in units of $\sigma(A_\textrm{lens})$. We run the estimators on coadded frequency channels (inverse-variance weighted \iac{``fcoadd''} for ACT, harmonic internal linear combination \iac{``HILC''} for SO). 
    \iac{\textit{Top}:} Reconstruction performed with the minimum variance MV estimator, \iac{with profile hardening}. We show the bias as a function of sky fraction \iac{(60\%: blue, 70\%: orange, 80\%: green)}. 
    Within the ACT \SI{70}{\percent} mask, the shift in $A_\textrm{lens}$ lies is below $0.3\sigma$ \iac{(vertical dashed gray line)}. For SO, however, it is unclear whether the bias remains at those levels, as not all models lie within this range. \iac{\textit{Bottom}:} Upper limit to the bias as a function of the temperature or polarization channels used in the reconstruction: both (MV) in \iac{blue}, temperature-only ($TT$) in \iac{orange} and polarization-only (MVPol) in \iac{green}. We use the 70\% mask and no profile hardening for this test. In general, we find much larger biases in temperature-only reconstructions in comparison to those using polarization channels.}
    \label{fig:alenstable_ilc}
\end{figure*}

For the ACT-like experiment, we perform an inverse-variance coaddition of the foreground maps at the different frequencies (``fcoadd''). We follow the same procedure as in ACT \citep{Frank_ACT_lensing_2024} to combine the maps from both \SI{90}{} and \SI{150}{\giga\hertz} frequency channels into a single ``coadded'' map $d_{\ell m}^{\textrm{coadd}}$. The coaddition is done in harmonic space,
\begin{equation}\label{eq:fcoadd}
    d^{\textrm{coadd}}_{\ell m} = \sum_f w_\ell^f d_{\ell m}^f, \quad f=\{\SI{90}{\giga\hertz}, \SI{150}{\giga\hertz}\},
\end{equation}
where the weights are derived from the real ACT noise at each frequency channel \citep{Atkins_2023} and are normalized to sum up to unity for every multipole $\ell$. We show in Figure \ref{fig:weights_night} the resulting weights for the temperature component of the foreground fields. The weights for the $E$ and $B$ components are derived from the same procedure, and follow the same trend: the \SI{90}{\giga\hertz} channel provides the largest contribution to the coadded map.\medskip

In the case of the SO-like experiment, we perform a Harmonic Internal Linear Combination \citep[HILC, e.g.][]{Bennett_2003, Eriksen_2004, Remazeilles_2011, McCarthy_2024}, similar to equation \eqref{eq:fcoadd} where the weights are now computed directly from the data covariance $\mathbb{C}$ (a $6\times 6$ matrix encoding the auto- and cross-power spectra at the SO frequencies):
\begin{equation}
    \bm{w} = \frac{\bm{1}^\top \mathbb{C}^{-1}}{\bm{1}^\top \mathbb{C}^{-1} \bm{1}},
\end{equation}
where both $\bm{w}$ and $\bm{1}$ are six-dimensional vectors (each entry corresponding to a different frequency channel)\footnote{We use the ILC method to isolate the CMB. We work in thermodynamic temperature units, for which the SED is constant across frequencies (c.f. equation \ref{eq:cmb_sed}), hence we have a six-dimensional unit vector.}. The resulting weights $\bm{w}$ are shown in Figure \ref{fig:sohilc_weights}. In the limit where the data are completely noise dominated, both ``fcoadd'' and HILC methods are equivalent. To construct the covariance, we use the auto- and cross-frequency temperature power spectra used in SO forecasting \citep{SO_2019}, and discussed in detail in the aforementioned paper. As a summary, they include lensed $TT$ CMB spectra, thermal and kinetic SZ effects, CIB and radio point source emission, in addition to synchrotron, free-free and anomalous microwave emission from \textit{Planck} \texttt{Commander} models \citep{Planck_commander2016} and thermal dust emission maps (``model 8'' \citep{Finkbeiner_1999} in \mbox{Sehgal} \textit{et al.} 2010 \cite{Sehgal_2010} simulations).
In polarization, we add to the lensed CMB mock power spectra that include SO-like noise\footnote{Generated using \url{https://github.com/simonsobs/V3_calc}.} and Galactic foreground emission. We measure the power spectra of PySM \texttt{d9} and \texttt{s4} models at \SI{353}{} and \SI{23}{\giga\hertz}, respectively. We extrapolate to the different SO channels using a modified black body spectrum for dust ($\beta_d=1.59$, $\Theta_d=19.6$; \citep{Planck_2015_dust}) and a simple black-body distribution for synchrotron ($\beta_s = -3.00$; \citep{Dunkley_2009, Planck_2016_low, Planck_2020_fgs}). As expected, we find that the weights are such that SO mid-frequency channels provide the dominant contribution to the HILC map across all multipoles.\medskip 

We present the results (with and without profile hardening) in Figure \ref{fig:raw4pt_dusts_GALS_HILC}. These correspond to reconstructions performed with the minimum-variance (MV) estimator. We report the resulting upper limits to the bias on the CMB power spectrum amplitude $A_\textrm{lens}$ in Figure \ref{fig:alenstable_ilc} (left panel). We find that within the ACT \SI{70}{\percent} mask, the shift in $A_\textrm{lens}$ lies below $0.3\sigma$ (dashed-gray vertical line). For SO, however, it is unclear whether the bias remains at those levels, as not all models lie within this range when the SO \SI{70}{\percent} mask is used. Finally, we find that extending the CMB analysis footprint to the \textit{Planck} \SI{80}{\percent} dust mask is potentially too risky if no further mitigation strategies are in place.\medskip

In Figure \ref{fig:alenstable_ilc} (right panel), we also quantify the bias on the lensing power spectrum amplitude for other estimators. We find that temperature-only reconstructions continue suffering from Galactic foreground biases. For ACT, this is expected as we are now including (although small, see Figure \ref{fig:weights_night}) contributions from the $\SI{150}{\giga\hertz}$ channel. For SO, we see that the HILC method performs very well for the PySM low-complexity model. We believe this due to the fact that the spectral parameters on this model remain uniform across the sky. The absence of decorrelation and the use of six frequency channels makes the HILC method ideal to remove the non-Gaussian Galactic contamination. However, we find larger biases with more complex models (reaching the $3\sigma$ level for Vansyngel \texttt{d1s1}). Additional foreground complexity, in addition to the fact that SO HILC relies much more on the $\SI{150}{\giga\hertz}$  frequency channel (see Figure \ref{fig:sohilc_weights}), is possibly the origin of these larger biases in comparison to the results obtained in \S\ref{sec:tonly}. In polarization, biases are smaller than in temperature. Therefore, quadratic estimators making use of polarization channels will be more robust against these biases. However, as we reach even lower noise levels and make use of iterative estimators, it is unclear what the impact of non-Gaussian Galactic foregrounds will be, and therefore future work is needed.\medskip

As mentioned earlier, the results presented here constitute an upper limit to the contamination produced by non-Gaussian Galactic foregrounds since we do not remove any of the expected lensing power spectrum biases. In Appendix \ref{sec:app_patches}, we explore a possible strategy to remove the well-known mean-field and Gaussian biases. The approach is based on a patch-by-patch reconstruction \iac{\citep[e.g.][]{Plaszczynski_2012}}, and had previously been proposed in \citep{Namikawa_2013, Challinor_2018}. However, we find this avenue unsuccessful and therefore choose not to include it when deriving our main results. \medskip

On the other hand, current analyses make use of the realization-dependent $N_L^{(0)}$ algorithm (RDN0; \citep{Namikawa_2013, Planck_lensing_2020}), which is insensitive to errors in the power spectra of the simulations used to compute this bias to first order in the fractional error. This algorithm is known to reduce off-diagonal $L$ correlations \iac{\citep{Nguyen_2019}}. In Appendix \ref{sec:app_rdn0}, we test whether it is successful at mitigating unaccounted non-Gaussian Galactic foreground biases with a connected 4-point function. In summary, we find that bias subtraction does not appear to qualitatively change the conclusions from our simpler upper limit, {with the caveat that those results are not fully converged due to limited simulation availability.}

\section{Conclusion}\label{sec:conclusion}

Weak gravitational lensing of the CMB is a mature observable for precision cosmology. However, a particular concern for CMB lensing studies might be Galactic foregrounds. These are inherently non-Gaussian and hence might mimic the characteristic signal that lensing estimators are designed to measure \citep{Challinor_2018, Beck_2020}. As a result, any unknown bias in the inferred lensing power spectrum amplitude can directly impact cosmological inferences.\medskip

In this work, we presented an analysis that quantifies the impact of Galactic dust non-Gaussianity on lensing reconstructions performed
with quadratic estimators in both temperature and polarization on the Atacama Cosmology Telescope (two frequency bands centered at \SI{90}{} and \SI{150}{\giga\hertz} with a white noise level of $\SI{12}{\micro\kelvin}_\mathrm{CMB}$-arcmin in temperature) and Simons Observatory (6 frequency channels and $\SI{6}{\micro\kelvin}_\mathrm{CMB}$-arcmin white-noise level in temperature) experiments.\medskip 

We applied the relevant quadratic estimators to a {comprehensive suite} of foreground simulations. The resulting measurement of the lensing power spectrum  $C_L^{\hat{\phi}\hat{\phi},\mathrm{fg}}$ constitutes an upper limit to the bias that Galactic non-Gaussian foregrounds pose to CMB lensing reconstruction measurements, and we used it to compute the bias on the inferred lensing power spectrum amplitude $A_\textrm{lens}$. The dust SED is such that we found biases in the lensing reconstruction almost 2 orders of magnitude larger in the 150 GHz channel than in the 90 GHz one. \medskip

In general, we found polarization measurements to present smaller biases than temperature ones. We also found, in line with previous works, that $\ell_{\textrm{min}}$ is the most sensitive to the presence of Galactic foregrounds, and recommend that all analyses continue ensuring consistency of different scale cuts. Biases also increased with sky fraction, and, although our results are conservative upper limits, to be cautious we advise against using a \textit{Planck} 80\% dust mask for CMB lensing reconstruction analyses without any further mitigation strategy in place.\medskip

After coadding the two frequency channels, and within the \SI{70}{\percent} mask region, Galactic foregrounds do not present a problem for the ACT experiment (the shift in $A_\textrm{lens}$ remains below 0.3$\sigma$). In SO, not all foreground models remain below this threshold after performing a HILC. Although our results are conservative upper limits, the wider area accessed by SO, together with the larger weight on the $\SI{150}{\giga\hertz}$ channel, suggest that further work on characterizing dust biases and determining the impact of mitigation methods is well motivated, especially for the largest sky fractions.\medskip

The non-Gaussian nature of Galactic foregrounds {poses a challenge to upcoming} CMB lensing analyses. This is relevant not only for precision cosmology with the CMB lensing power spectrum, but it also impacts other probes such as the detection of inflationary \emph{B}-modes \citep{Namikawa_2023,Namikawa_2025b} or post-Born lensing curl \emph{B}-modes \citep{Robertson_2023, Robertson_2025}. This work was an {early} analysis to quantify the level of contamination from Galactic emission on current ground-based CMB experiments, and motivates further investigation to develop mitigation strategies for this foreground. In particular, ongoing work is studying the impact of component separation methods that use positional information, such as wavelets or needlets \citep[see e.g.][]{Coulton_2024, Goldstein_2025}, on the Galactic foreground problem for lensing. {In addition, synergies with upcoming space telescopes such as the LiteBIRD satellite \citep{Litebird_2020, LiteBird_2023} will better inform our component separation strategy.} \medskip

\begin{acknowledgements}

We thank Toshiya Namikawa and Anthony Challinor for helpful input and discussions, Zach Atkins for providing the core code to plot masks in CAR projection.
IAC acknowledges support from Fundaci\'on Mauricio y Carlota Botton and the Cambridge International Trust. CHC acknowledges funding by Agencia Nacional de Investigaci\'on y Desarrollo, Fondo Nacional de Desarrollo Cient\'ifico y Tecnol\'ogico Postdoc fellowship 3220255 and by Agencia Nacional de Investigaci\'on y Desarrollo, BASAL, FB210003. BDS acknowledges support from the European Research Council (ERC) under the European Union’s Horizon 2020 research and innovation programme (Grant agreement No. 851274). NS acknowledges support from DOE award number DE-SC0025309. MM acknowledges support from NSF grants AST-2307727 and  AST-2153201 and NASA grant 21-ATP21-0145. OD acknowledges support from a SNSF Eccellenza Professorial Fellowship (No. 186879). JCH acknowledges support from the Sloan Foundation and the Simons Foundation. \medskip

This research has also made extensive use of \textsc{orphics}\footnote{\url{https://github.com/msyriac/orphics}} and the \textsc{astropy} \citep{astropy}, \textsc{numpy} \citep{numpy}, \textsc{scipy} \citep{scipy} packages. We also acknowledge use of the \textsc{matplotlib} \citep{matplotlib} package to produce the plots in this paper. 
This work was performed using resources provided by the Cambridge Service for Data Driven Discovery (CSD3) operated by the University of Cambridge Research Computing Service (\url{www.csd3.cam.ac.uk}), provided by Dell EMC and Intel using Tier-2 funding from the Engineering and Physical Sciences Research Council (capital grant EP/T022159/1), and DiRAC funding from the Science and Technology Facilities Council (\url{www.dirac.ac.uk}). In particular, this work used the DiRAC Data Intensive service. The DiRAC component of CSD3 at Cambridge was funded by BEIS, UKRI and STFC capital funding and STFC operations grants. DiRAC is part of the UKRI Digital Research Infrastructure. This is not an official Simons Observatory Collaboration paper.\medskip

\end{acknowledgements}

\section*{Data Availability}

The \textit{Planck} \SI{353}{\giga\hertz} map and analysis masks were accessed from the \textit{Planck}\footnote{\textit{Planck}~(\url{https://www.esa.int/Planck}) is an ESA science mission with instruments and contributions directly funded by ESA Member States, NASA, and Canada.}~Legacy Archive at \url{https: //pla.esac.esa.int/}. All other masks are available on NERSC upon request. The dust filament maps were generated using the \textsc{dustfilaments} \citep{DF_2022, Hervias_2024} simulation and will be shared on reasonable request to the corresponding author. The code to generate all other data is available at \url{https://github.com/iabrilcabezas/dustbias}. 

\bibliographystyle{act_titles} 
\bibliography{main}

\appendix

\section{A pedagogical introduction to quadratic estimators in CMB lensing}\label{sec:app_QE}

\begin{figure}
\includegraphics[width = 0.99\columnwidth]{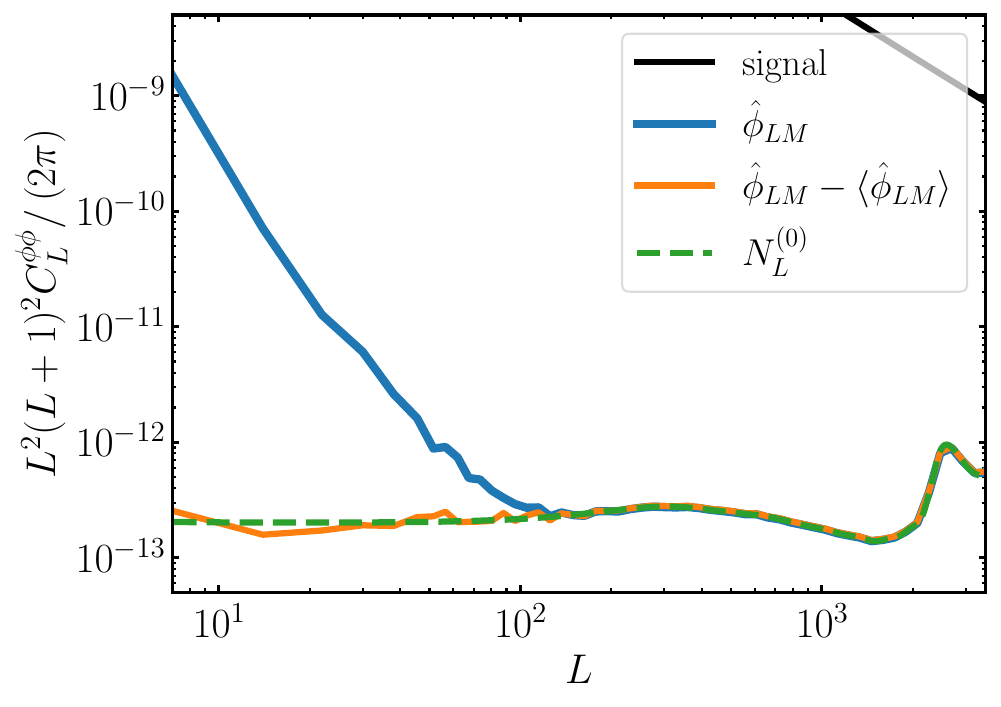}
\caption{Pedagogical introduction to CMB lensing biases. We run the $TT$ quadratic estimator on a masked Gaussian (unlensed) field. The recovered signal (solid blue) is equivalent to the reconstruction noise on small scales, which can be computed analytically and is shown as a dashed green line. The mask is also a source of statistical anisotropy and generates the so-called mean-field bias. This bias can be computed from simulations and then subtracted from the measured signal (solid orange). See text for further details.}\label{fig:explain_biases_wgauss} 
\end{figure}

\begin{figure*}
\includegraphics[width=0.9\textwidth]{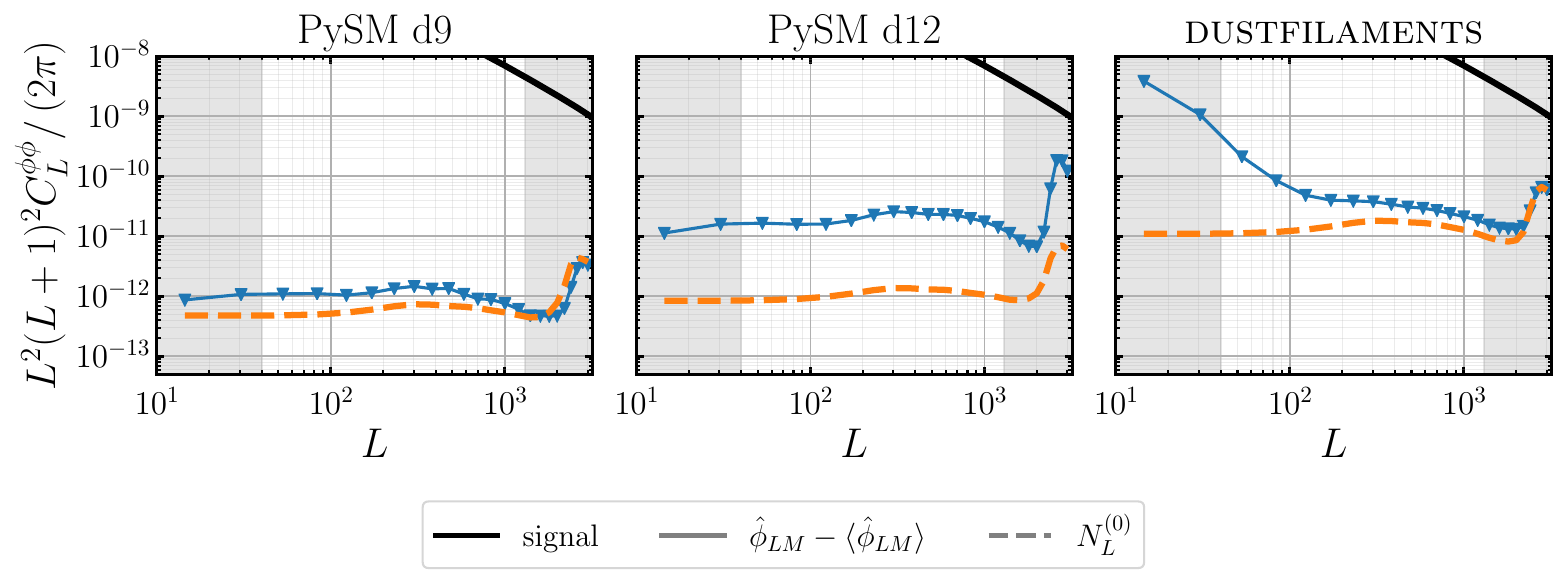}
\caption{Lensing reconstruction performed on different dust models, where we divide the footprint in different $\SI{50}{\deg\squared}$ patches, perform the lensing reconstruction locally, and then sum over different patches. With this approach, we can estimate the mean-field and $N_L^{(0)}$ bias (dashed orange line),   although with varying levels of success. The results here correspond to the $\SI{90}{\giga\hertz}$ ACT channel on the ACT $60\%$ mask, for the temperature-only estimator including profile hardening.}\label{fig:patches} 
\end{figure*} 

For simplicity, we continue the discussion working in the flat-sky approximation and focusing only on temperature fields. Lensing breaks translation invariance in the CMB. As described in the main text, when averaging over all possible unlensed CMB realizations, different modes in the CMB now become correlated due to lensing (equation \ref{eq:TT}). That expression suggests the following estimator for $\phi(\mathbf{L})$:
\begin{equation}
    \hat{\phi}(\mathbf{L}) \sim \frac{T\left(\mathbf{l}\right)T(\mathbf{L}-\mathbf{l})}{f(\mathbf{l}, \mathbf{L})},\end{equation}
where we can combine all different pairs of modes applying inverse-variance weighting:\footnote{Given a set of measurements $\{ \hat{X}_i \}$, they can be combined into a single estimate for $X$ as $\hat{X} = \sum_i w_i \hat{X}_i$. The weights that provide an unbiased estimate for $\hat{X}$ (expressed with the condition $\sum_i w_i=1$) while minimizing its variance are $w_k = \left(\sum_i 1/ \sigma_i^2\right)^{-1} \cdot 1/\sigma_k^2 $, where $\sigma_i^2$ is the variance of each measurement.}
\begin{equation}\label{eq:hatpiraw}
     \hat{\phi}(\mathbf{L})  = A_\mathbf{L}^\phi \int \frac{\mathrm{d}^2\mathbf{l}}{\left(2\pi\right)^2} \frac{1}{\mathrm{Var}\left(\frac{T\left(\mathbf{l}\right)T(\mathbf{L}-\mathbf{l})}{f(\mathbf{l}, \mathbf{L})}\right)} \frac{T\left(\mathbf{l}\right)T(\mathbf{L}-\mathbf{l})}{f(\mathbf{l}, \mathbf{L})},
\end{equation}
where $A_\mathbf{L}^{\phi}$ is inversely proportional to the total sum of inverse-variance weights,
\begin{equation}\label{eq:araw}
      A_\mathbf{L}^\phi  =  \left[\int \frac{\mathrm{d}^2\mathbf{l}}{\left(2\pi\right)^2} \frac{1}{\mathrm{Var}\left(\frac{T\left(\mathbf{l}\right)T(\mathbf{L}-\mathbf{l})}{f(\mathbf{l}, \mathbf{L})}\right)}\right]^{-1}.
\end{equation}
Using Wick's theorem, 
\begin{equation}\label{eq:vart}
    \mathrm{Var}\bigl(T\left(\mathbf{l}\right)T\left(\mathbf{L}-\mathbf{l}\right)\bigr)= 2 C_\ell^{\mathrm{obs}} C_{| \mathbf{L} - \mathbf{l}|}^{\mathrm{obs}}, 
\end{equation}
where $C_\ell, C_{| \mathbf{L} - \mathbf{l}|}$ are the observed (lensed) CMB spectra including noise. Substituting this expression in equations \eqref{eq:hatpiraw}, \eqref{eq:araw}, we obtain equations \eqref{eq:phihat} and \eqref{eq:anorm}, respectively.\medskip

When the temperature fields are combined using inverse-variance weights computed from the total power in the temperature map, as it is the case in equation \eqref{eq:phihat}, the weights are so-called optimal. However, if those same weights, or filters, are used on a different temperature map (e.g. a foreground map), those weights will no longer be optimal. In that case, the variance of the reconstructed field will be given by\footnote{This problem is equivalent to computing $\mathrm{Var}(\hat{X})$, 
\begin{equation}
    \mathrm{Var}(\hat{X}) = \mathrm{Var}\left(A \sum_i w_i \hat{X}_i\right) = A^2 \sum_i w_i^2 \mathrm{Var}(\hat{X}_i)
\end{equation}
where $w_i$ are derived from the total (observed) CMB map (equation \ref{eq:vart}), whereas $\mathrm{Var}(\hat{X}_i)$ is proportional to the total foreground power.} \citep{Hu_2002, MacCrann_2024}: 
\begin{equation}\label{eq:n0_fg}
    N^{(0)}_L =  \bigl(A_\mathbf{L}^\phi\bigr)^2 \int \frac{\mathrm{d}^2\mathbf{l}}{\left(2\pi\right)^2} \frac{f^2(\mathbf{l}, \mathbf{L}) C_\ell^{\mathrm{fg}} C_{| \mathbf{L} - \mathbf{l}|}^\mathrm{fg}}{2 \left(C_\ell^{\mathrm{obs}} C_{| \mathbf{L} - \mathbf{l}|}^{\mathrm{obs}}\right)^2}.
\end{equation}

The normalization of the quadratic estimator is equivalent to the lensing power spectrum reconstruction noise. This is clear when running the lensing reconstruction on a Gaussian field. In this case, no lensing signal should be present. The output of the reconstruction is simply the reconstruction noise, sourced by the disconnected four-point function of the field. This is pictured in Figure \ref{fig:explain_biases_wgauss}, where we run the temperature quadratic estimator on a Gaussian field generated from a given power spectrum. The measured signal (solid blue) perfectly agrees with the analytical expectation for the $N_L^{(0)}$ bias (dashed green) at high $L$. At low $L$, there is the so-called mean-field bias, which is due to the mask. The mask is a source of statistical anisotropy, present even in the absence of lensing. This is picked up by the quadratic estimator. In order to remove this contribution, we run the estimator on several masked Gaussian simulations with the same power spectrum as our fiducial map, and take the average $\langle \hat{\phi}_{LM}\rangle$. We remove the mean-field bias $\langle \hat{\phi}_{LM}\rangle$ from the reconstructed $\hat{\phi}_{LM}$ before taking the power spectrum. The result is shown as the solid orange line in Figure \ref{fig:explain_biases_wgauss}. With this correction, the reconstructed signal agrees with the Gaussian bias on all scales.

\section{Local lensing reconstructions}\label{sec:app_patches}

{To fully capture the impact of non-Gaussian dust on lensing spectrum measurements, the steps of mean field subtraction and realization-dependent bias subtraction would need to be simulated accurately. However, this is challenging because to accurately capture these steps, CMB signal and instrument noise would need to be added to the dust simulations, which would make it difficult to obtain converged estimates of very small dust biases. Despite these challenges, we will attempt this approach in the next appendix.} \medskip

{In this appendix, however, we aim to obtain a simple, approximate (and more converged) estimate of the impact of mean-field and $N_L^{(0)}$ bias subtraction.} {Note that simply using the mean dust two-point function in the map to compute these mean-field and bias terms is not sufficient, since this does not capture position-dependence in the dust power; correspondingly, we find that this approach underestimates the bias present in the reconstruction for all models beyond a simple Gaussian dust field.}\medskip

The results provided in the main text correspond to upper limits to the total bias on the CMB lensing power spectrum amplitude due to the presence of non-Gaussian Galactic foregrounds. We found that the bias should lie below percent level. We confirm this result by running the full ACT lensing pipeline on two different simulations, which include a non-Gaussian dust realization from \textsc{dustfilaments} \citep{DF_2022} and PySM \texttt{d12}, respectively, using the MV estimator.\medskip

A possible workaround is to divide the footprint in different patches, perform the lensing reconstruction locally, and then sum over all different patches \iac{\citep{Plaszczynski_2012, Namikawa_2013, Challinor_2018}}. {This is an approximation to the performance of the realization-dependent bias subtraction algorithm, making our results less conservative (and perhaps overly optimistic) when compared with the ones presented in the main text.} With this approach, the two-point function within each patch will represent the power within each patch more closely. We perform this exercise for the profile-hardened temperature-only estimator at $\SI{90}{\giga\hertz}$ on the ACT 60\% mask and noise levels. We choose individual patches of area $\SI{50}{\deg\squared}$. \medskip 

\begin{table}
\renewcommand{\arraystretch}{1.7} 
\centering
\caption{Upper limit to the bias of the lensing power spectrum amplitude for different dust models in a configuration similar to the ACT experiment in the $60\%$ mask for the $\SI{90}{\giga\hertz}$ channel. We report results obtained for the $TT$ estimator with profile hardening. We compare performing the reconstruction locally and then summing over the different patches (``patches'') versus simply performing the reconstruction on the full footprint without any bias subtraction (``full-sky''). The former allows to compute estimates for the mean-field and $N_L^{(0)}$ biases, which we subtract. In general, this approach allows for a reduction of the bias but it is not perfect, as can be seen in the accompanying Figure \ref{fig:patches}.\vspace{2mm}}
\begin{tabular}{|>{\centering\arraybackslash}p{2cm}|>{\centering\arraybackslash}p{2.5cm}|>{\centering\arraybackslash}p{2.5cm}|}\hline
$100\times\Delta A_\textrm{lens}$ & Patches & Full-sky \\ \hline
{PySM \texttt{d9}} & 0.002 (0.001$\sigma$) & 0.003 (0.002$\sigma$) \\ \hline
{PySM \texttt{d12}} & 0.084 (0.054$\sigma$) & 0.081 (0.053$\sigma$) \\ \hline
{\textsc{dustfil}.} & 0.067 (0.043$\sigma$) & 0.120 (0.078$\sigma$) \\ \hline
\end{tabular}
\label{table:patches_act}
\end{table}

In Figure \ref{fig:patches}, we show our estimate for the $N_L^{(0)}$ bias computed this way (following equation \ref{eq:n0_fg}) for three different models (dashed lines). We compare this to the power spectrum of the reconstructed $\hat{\phi}_{LM}$, where we have previously removed an estimate of the mean-field bias within each patch following the approach described in Appendix \ref{sec:app_QE} (solid lines).  We find that this approach is the most successful with the \textsc{dustfilaments} model. For PySM \texttt{d9}, the $N_L^{(0)}$ bias is slightly overestimated, which leads to a negative lensing power spectrum at high $L$. In the case of PySM \texttt{d12}, the $N_L^{(0)}$ bias is greatly underestimated. We believe this is due to the presence of localized dust blobs in some patches, which makes the 2-point function not representative of the power across the patch even for a reduced sky area. These results are also captured in Table \ref{table:patches_act}, where we derive upper limits to the shift induced on the lensing power spectrum amplitude with this approach. In the next section, we bypass the aforementioned issues by performing the full reconstruction (including bias subtraction) with the machinery of modern lensing analyses.\medskip

\section{RDN0}\label{sec:app_rdn0}

\begin{figure}
\includegraphics[width = 0.95\columnwidth]
{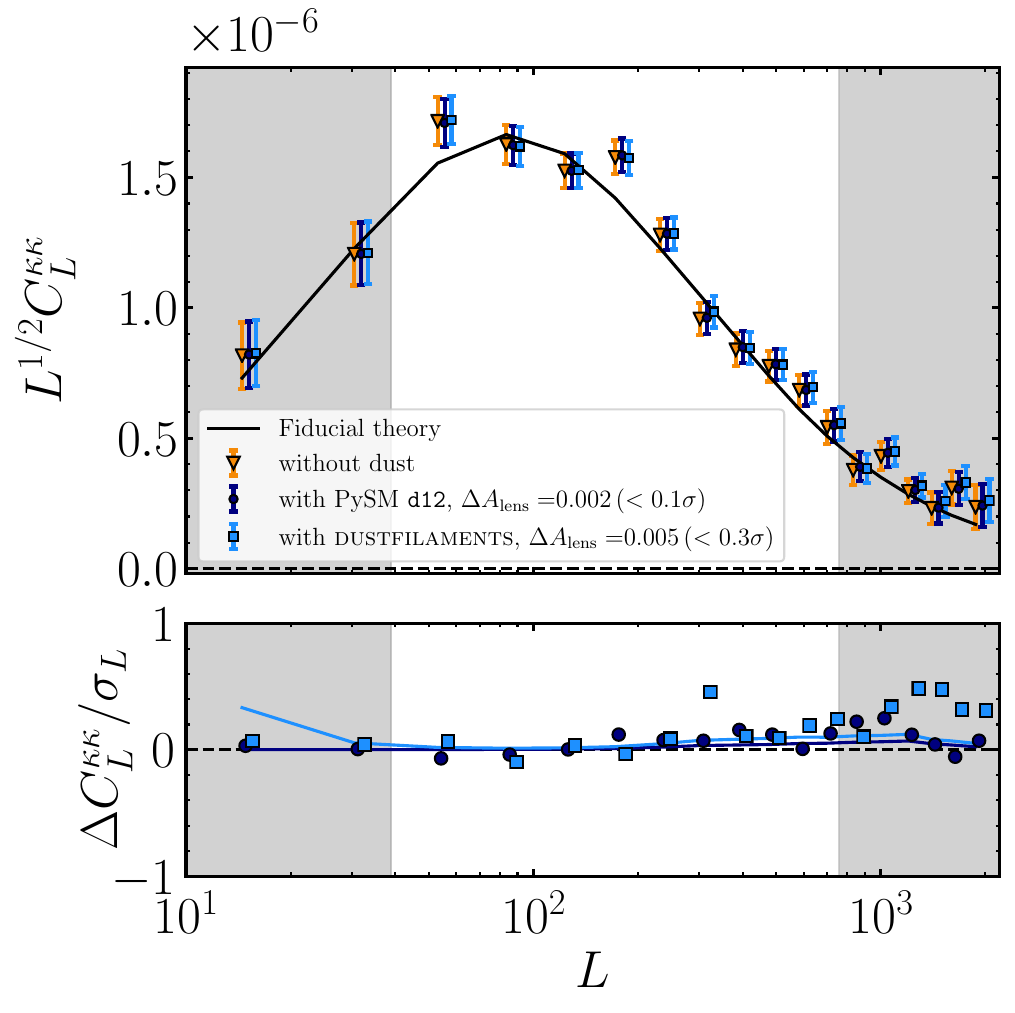}
\includegraphics[width = 0.9\columnwidth]{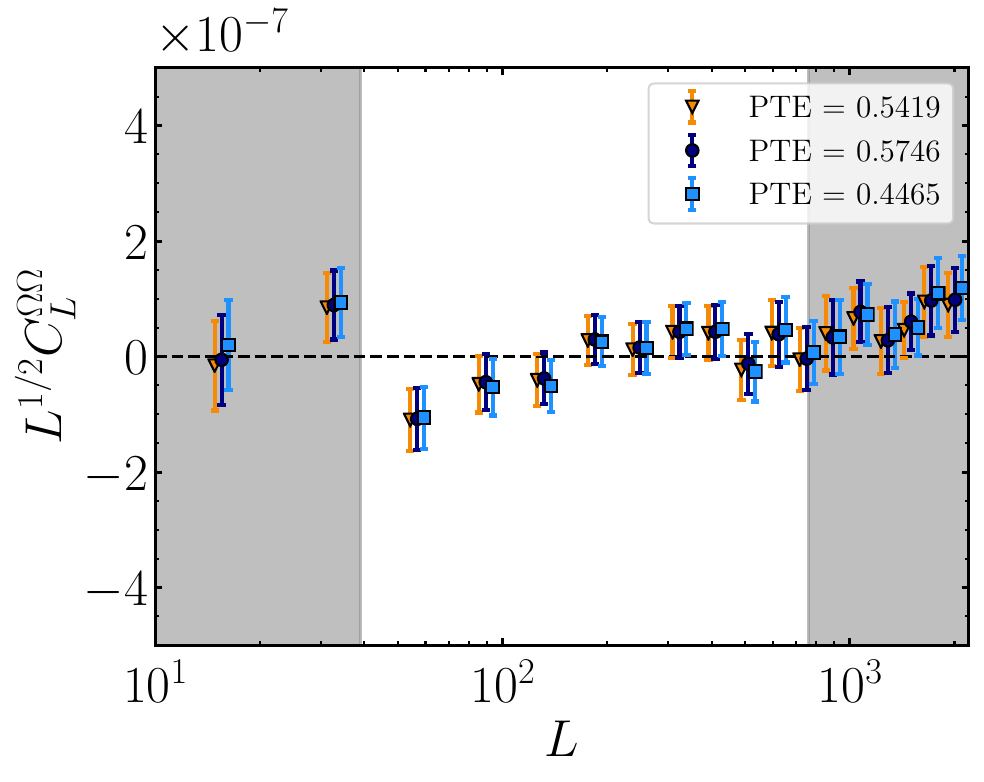}
\caption{We run the full ACT lensing pipeline on two simulations that include a non-Gaussian dust realization: \textsc{dustfilaments} (light blue squares) and PySM \texttt{d12} (dark blue circles). We compare them to a reconstruction done on a simulation that does not include dust (orange triangles). The results shown correspond to the MV estimator. \textit{Top}: We confirm that the bias on the CMB lensing power spectrum amplitude due to the presence of a non-Gaussian dust field is consistent with producing {a negligible shift in $A_\textrm{lens}$ in comparison to its measurement uncertainty $\sigma$}, and in agreement with the results found on the main text (solid lines). \textit{Bottom}: The curl component of the reconstruction is compatible with zero. We are not sensitive to any dust bias.}\label{fig:rdn0} 
\end{figure}

We show the reconstructed bandpowers in the top panel of Figure \ref{fig:rdn0}. Note that both simulations include a map-based realistic noise realization computed with \textsc{mnms} \citep{Atkins_2023}, and the reconstruction is done employing the cross-correlation-based quadratic estimator \citep{Madhavacheril_2020} which is immune to noise bias. Moreover, we subtract all relevant biases following the same procedure as \citep{Frank_ACT_lensing_2024}. This includes the realization-dependent $N_L^{(0)}$ algorithm, which absorbs small changes in the data due to the presence of dust. We estimate the lensing amplitude parameter $A_\textrm{lens}$ by fitting the measured bandpowers to the baseline cosmology prediction. In the range $40<L<763$, and find:
\begin{align*}
\textrm{No dust}:\  &    A_\textrm{lens} =  1.021 \pm 0.021,\ \\ %
\textsc{dustfilaments}: \ & A_\textrm{lens}^{\textrm{fg}} =  1.026 \pm 0.021,\ \\ 
\textrm{PySM}~\texttt{d12}:\  &    A_\textrm{lens}^{\textrm{fg}} =  1.023 \pm 0.021.\ \\ 
\end{align*}
{In all cases, the shift in $A_\textrm{lens}$ is less than $0.3\sigma$, where $\sigma$ is the measurement error on $A_\textrm{lens}$}. Within the extended range $40<L<1300$, we find a slight increase in the bias, at the $0.4\sigma$ level:
\begin{align*}
\textrm{No dust}:\  &    A_\textrm{lens} =  1.023 \pm 0.020,\ \\ 
\textsc{dustfilaments}: \ & A_\textrm{lens}^{\textrm{fg}} =  1.031 \pm 0.020, \ \\ 
\textrm{PySM}~\texttt{d12}:\  &    A_\textrm{lens}^{\textrm{fg}} =  1.027 \pm 0.020.\ 
\end{align*}
This bound is in agreement with what we found in the main text (\S\ref{sec:ilc}, Figures \ref{fig:raw4pt_dusts_GALS_HILC}--\ref{fig:alenstable_ilc}), where we also reported an upper limit for the bias of $0.3\sigma-0.4\sigma$ for these models. Finally, in the bottom panel of Figure \ref{fig:rdn0} we show the curl component of both reconstructions,  {where we also compute the probability to exceed (PTE) with respect to a zero curl. No curl is significantly detected.} \medskip


\section{Lensing reconstruction is not sensitive to $EB$ power spectra}\label{sec:app_eb}

\begin{figure}[t]
\vspace{7mm}
\includegraphics[width = 0.9\columnwidth]{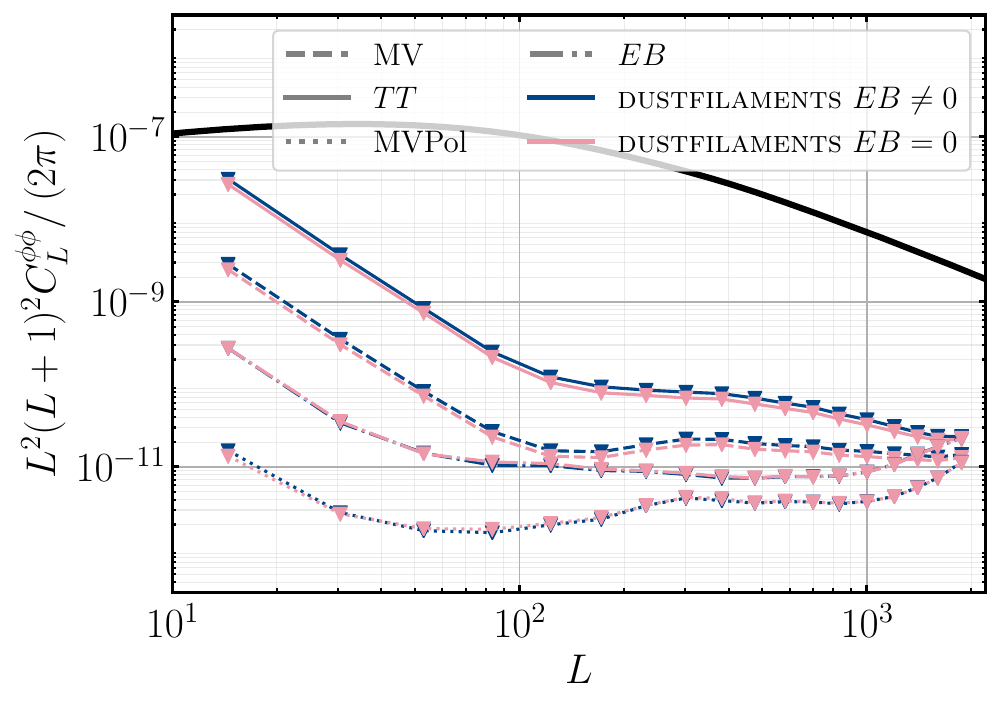}
\caption{The absence of $EB$ correlation on PySM models \citep{pysm} is not the cause for ACT/SO experiments to be less sensitive to non-Gaussian Galactic foreground biases in polarization. Here we show the difference in the reconstruction on \textsc{dustfilaments} \citep{DF_2022} simulations that include and do not include $EB$ correlations \citep{Hervias_2024}, for a SO-like configuration at  \SI{93}{\giga\hertz} on the SO 70\% mask. We find that the level of $EB$ correlations compatible with existing data are too small to have any detectable contribution to the trispectrum. The small differences seen in temperature are due to the way these simulations are constructed.}\label{fig:ebcorr} 
\end{figure}

In Figure \ref{fig:ebcorr}, we see that estimators that make use of polarization measurements are less biased. This is not a result of the lack of $EB$ correlation present on these models. Any non-zero $EB$ signal cannot produce a mean-field bias in the $EB$ estimator. However, it could generate some non-zero bias at the trispectrum level, as pointed out by \citep{Beck_2020}. In Figure \ref{fig:ebcorr}, we show that the level of $EB$ correlations compatible with existing data are too small to have any detectable contribution to the trispectrum.\medskip

We run the quadratic estimator on two sets of \textsc{dustfilaments} simulations, one of which includes $EB$ correlations \citep{Hervias_2024} at the level compatible with existing data \citep{planck_int_XXX, Planck_2020_XI}. We only find differences in the temperature-only reconstructions (that are then propagated into the MV estimator). This is due to small differences in the temperature component of the two simulations, present in order to achieve the non-zero parity-violating correlation: the orientation of the filaments with respect to the underlying magnetic field is manipulated, rendering populations of filaments with different orientations and therefore slightly different temperature fields \citep{Hervias_2024}.\medskip

\section{Extension of figures shown in the main text}\label{ap:full_versions}

\iac{We show extended versions of Figures   \ref{fig:raw4pt_TT_exp_gals} and \ref{fig:raw4pt_dusts_GALS_HILC}, in order to showcase the results for the wider variety of foreground models studied in this work.\medskip}

\begin{figure*}
\includegraphics[width = 0.74\textwidth]{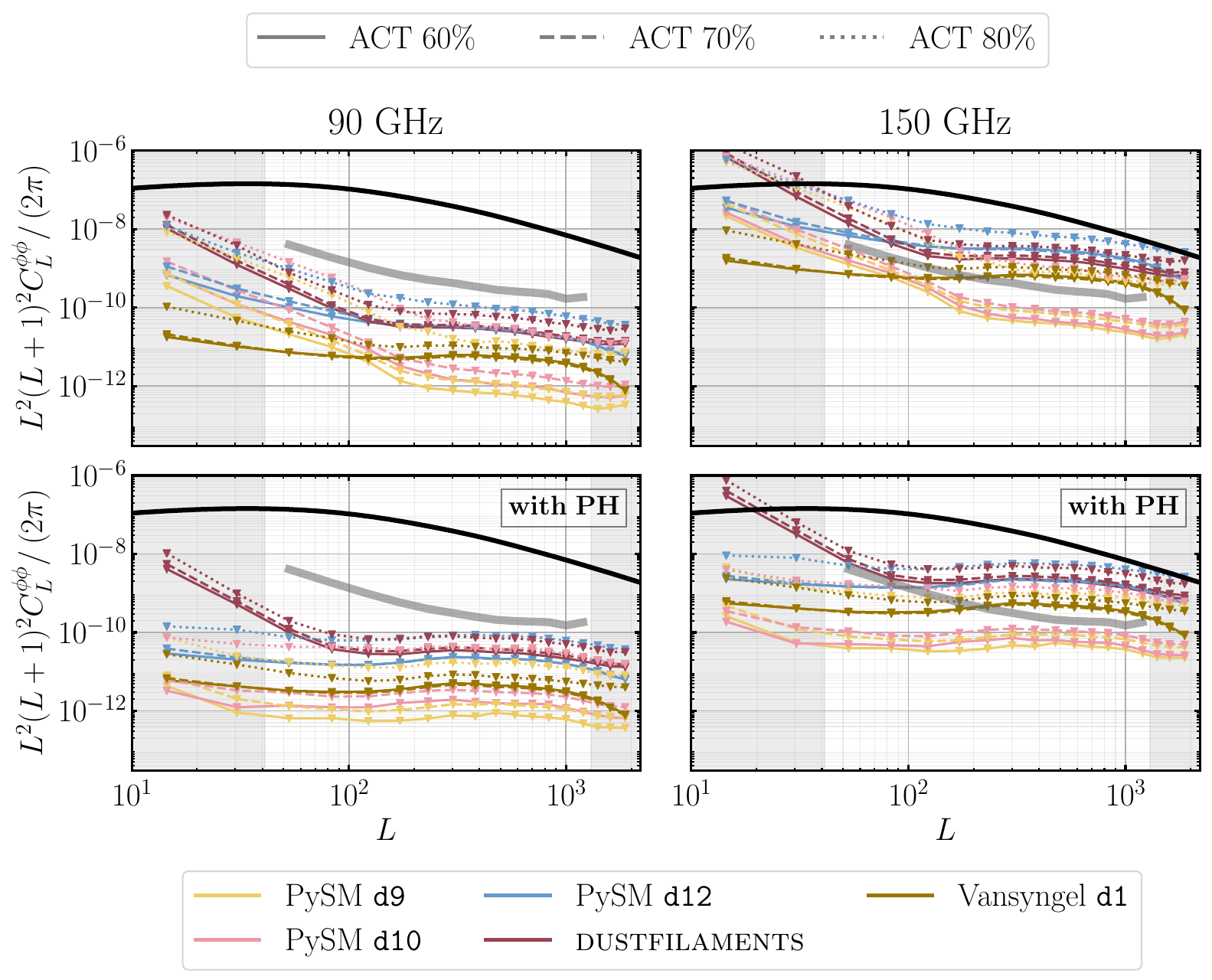}
\includegraphics[width = 0.74\textwidth]{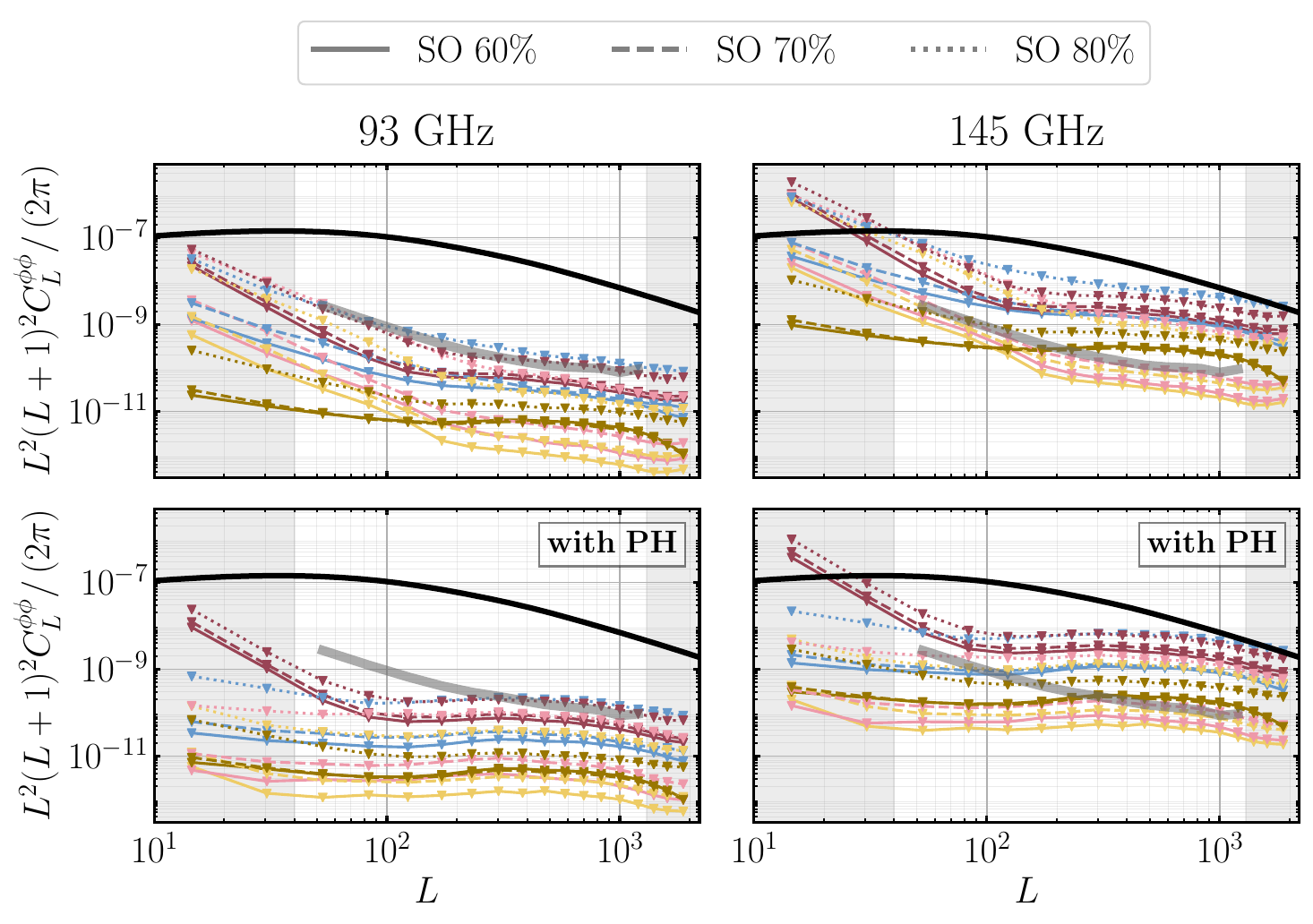}
\caption{Upper limits to the level of Galactic dust contamination on CMB lensing reconstruction measurements in temperature, for a variety of foreground models studied in this work \iac{(extension of Figure \ref{fig:raw4pt_TT_exp_gals})}.
}\label{fig:raw4pt_TT_exp_gals_full} 
\end{figure*}

\begin{figure*}
\includegraphics[width = 0.95\textwidth]{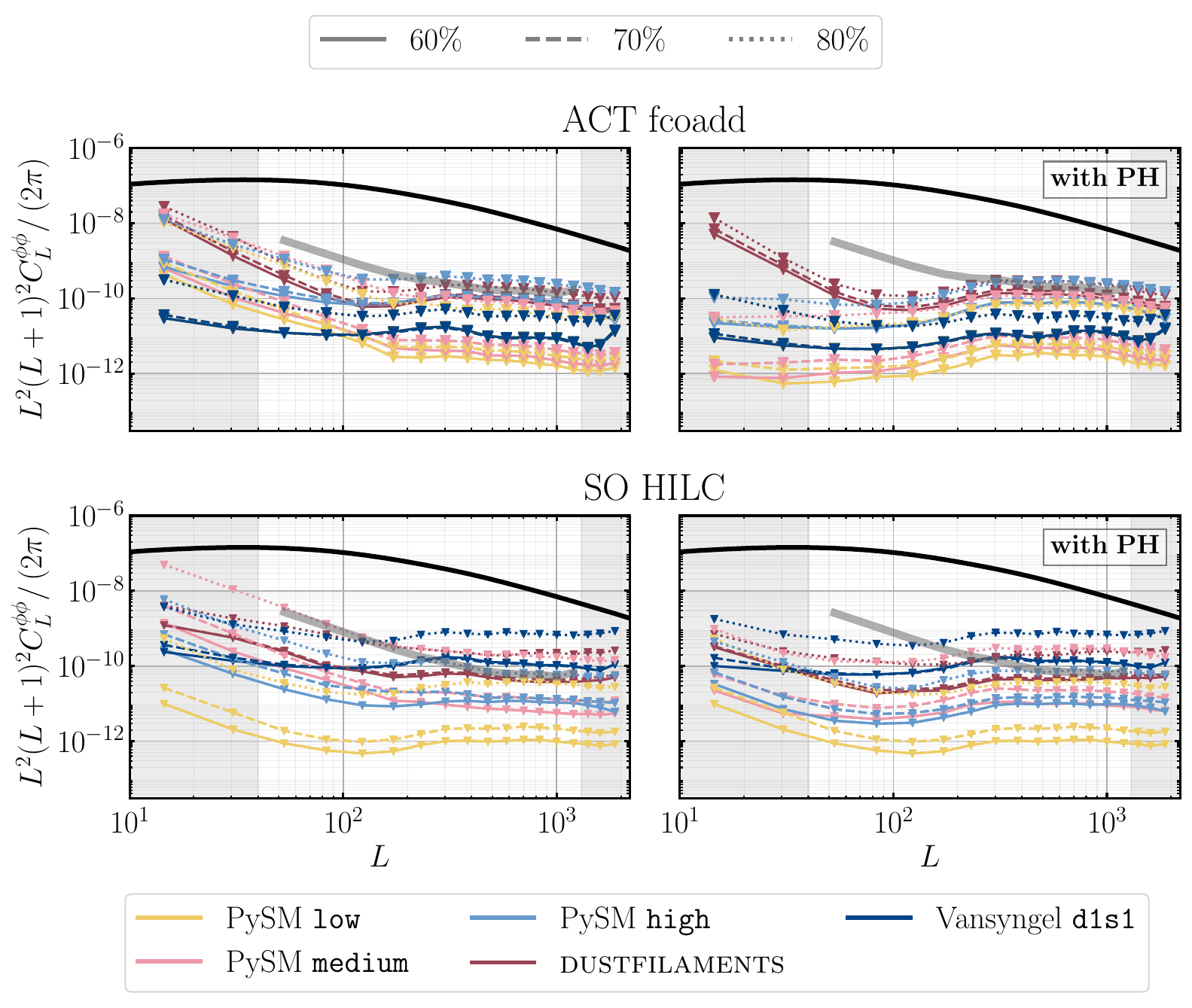}
\caption{Upper limits on the amplitude of the non-Gaussian Galactic foreground bias using the minimum-variance quadratic estimator within ACT and SO, for different foreground models and sky fractions \iac{(extension of Figure \ref{fig:raw4pt_dusts_GALS_HILC})}.
}\label{fig:raw4pt_dusts_GALS_HILC_full} 
\end{figure*}

\end{document}